\documentclass[5p,sort&compress]{elsarticle}
\usepackage{amssymb, amsmath,color, hyperref, graphicx}
\usepackage[switch]{lineno}
\begin{document}
\newcommand {\todo}[1] {\textcolor{red}{#1}}

\newcommand {\m}[1] {\ensuremath{\mu_{#1}}}
\newcommand {\tc}[1] {\ensuremath{T_c(\mu_{#1})}}
\newcommand {\tco} {\ensuremath{T_c(0)}}
\newcommand {\kt}[1] {\ensuremath{\kappa_2^{#1}}}
\newcommand {\kf}[1] {\ensuremath{\kappa_4^{#1}}}
\newcommand {\op}{\ensuremath{\Sigma}}
\newcommand {\qbq}[1] {\ensuremath{\bar{#1}{#1}}}
\newcommand {\sust}{\ensuremath{\chi^\op}}
\newcommand {\sus} {\ensuremath{\chi}}
\newcommand {\opc}[1] {\ensuremath{C^\op_{#1}}}
\newcommand {\susc}[1] {\ensuremath{C^\sus_{#1}}}
\newcommand {\ct}{\ensuremath{T_c^0}}
\newcommand {\Done} {\ensuremath{T\partial_T}}
\newcommand {\D}[1] {\ensuremath{T^{#1}\partial_T^{#1}}}
\newcommand {\nt} {\ensuremath{N_\tau}}


\begin{frontmatter}
\title{Chiral crossover in QCD at zero and non-zero chemical potentials \tnoteref{t1}}
\tnotetext[t1]{\textbf {HotQCD Collaboration}}


\author[1]{A.\ Bazavov}
\author[2]{H.-T.\ Ding}
\author[3]{P.\ Hegde}
\author[2,4]{O.\ Kaczmarek}
\author[4,5]{F.\ Karsch}
\author[5]{N.\ Karthik}
\author[4]{E.\ Laermann\fnref{a1}}
\author[4]{Anirban\ Lahiri}
\author[5]{\hspace{3em}R.\ Larsen}
\author[2]{S.-T.\ Li }
\author[5]{Swagato\ Mukherjee}
\author[6]{H.\ Ohno}
\author[5]{P.\ Petreczky}
\author[4]{H.\ Sandmeyer}
\author[4]{C.\ Schmidt}
\author[7]{S.\ Sharma}
\author[5]{\hspace{3em}P.\ Steinbrecher}

\address[1]{Department of Computational Mathematics, Science and Engineering and
Department of Physics and Astronomy,\\ Michigan State University, East Lansing, MI
48824, USA}
\address[2]{Key Laboratory of Quark \& Lepton Physics (MOE) and Institute of
Particle Physics,\\ Central China Normal University, Wuhan 430079, China}
\address[3]{Center for High Energy Physics, Indian Institute of Science,
Bangaluru 560012, India}
\address[4]{Fakult\"at f\"ur Physik, Universit\"at Bielefeld, D-33615 Bielefeld,
Germany}
\address[5]{Physics Department, Brookhaven National Laboratory, Upton, NY 11973,
USA}
\address[6]{Center for Computational Sciences, University of Tsukuba, Tsukuba,
Ibaraki 305-8577, Japan}
\address[7]{Department of Theoretical Physics, The Institute of Mathematical
Sciences, Chennai 600113, India}

\fntext[a1]{deceased}

\begin{abstract}

We present results for pseudo-critical temperatures of QCD chiral crossovers at zero
and non-zero values of baryon (\(B\)), strangeness (\(S\)), electric charge (\(Q\)),
and isospin (\(I\)) chemical potentials \m{X=B,Q,S,I}. The results were obtained
using lattice QCD calculations carried out with two degenerate up and down dynamical
quarks and a dynamical strange quark, with quark masses corresponding to physical
values of pion and kaon masses in the continuum limit.  By parameterizing
pseudo-critical temperatures as \( \tc{X} = \tco \left[ 1 -\kt{X}(\m{X}/\tco)^2
-\kf{X}(\m{X}/\tco)^4 \right] \), we determined \kt{X} and \kf{X} from Taylor
expansions of chiral observables in \m{X}. We obtained a precise result for
\(\tco=(156.5\pm1.5)\)~MeV. For analogous thermal conditions at the chemical
freeze-out of relativistic heavy-ion collisions, \emph{i.e.}, \(\m{S}(T,\m{B})\) and
\(\m{Q}(T,\m{B})\) fixed from strangeness-neutrality and isospin-imbalance, we found
\kt{B}=0.012(4)\ and \kf{B}=0.000(4).  For \(\m{B}\lesssim300\)~MeV, the chemical
freeze-out takes place in the vicinity of the QCD phase boundary, which coincides
with the lines of constant energy density of \(0.42(6)~\mathrm{GeV/fm}^3\) and
constant entropy density of \(3.7(5)~\mathrm{fm}^{-3}\).

\end{abstract}
\end{frontmatter}
\section{Introduction}\label{sc:intro}

The spontaneous breaking of the chiral symmetry in quantum chromodynamics (QCD)
is a key ingredient for explaining the masses of hadrons that constitute almost the
entire mass of our visible Universe. Lattice-regularized QCD calculations have
demonstrated (near) restoration of the broken chiral symmetry in QCD at high
temperature (\(T\)) through a smooth crossover~\cite{Ding:2015ona}. The chiral
crossover temperature of QCD  marks the epoch at which massive hadrons were born
during the evolution of the early Universe. The chiral crossover in the early
Universe took place at vanishingly small baryon chemical potential \(\m{B}\), although
the electric charge chemical potential \(\m{Q}\) at that stage might have been
non-vanishing~\cite{Wygas:2018otj}. For \(\m{B}>0\), \emph{i.e.}, when QCD-matter is
doped with an excess of quarks over antiquarks, the chiral crossover in QCD might
lead to a rich phase diagram in the \(T\)-\(\m{B}\) plane~\cite{Fukushima:2010bq}.
The phase structure of QCD-matter in the \(T\)-\(\m{B}\) plane can be probed in
various ongoing and upcoming relativistic heavy-ion collision
experiments~\cite{Busza:2018rrf}. The phase diagram of QCD can be explored in these
experiments if the so-called chemical freeze-out takes place in the proximity of the
chiral crossover phase boundary in the \(T\)-\(\m{B}\) plane~\cite{Andronic:2017pug}.
Since the colliding heavy-ions do not carry any net strangeness, the medium formed in
the process is strangeness-neutral, \emph{i.e.}, characterized by \(n_S=0\), \(n_S\)
being the net strangeness-density. Additionally, the proton-to-neutron ratio of the
colliding nuclei determines the ratio of net charge-density (\(n_Q\)) to net
baryon-density (\(n_B\)) of the produced medium. For the most common relativistic
heavy-ion collisions with Au+Au and Pb+Pb this ratio turns out to be \(n_Q/n_B=0.4\);
consequently, the corresponding chemical freeze-out stages also respect the
conditions \(n_S=0\) and \(n_Q=0.4n_B\).

With the aid of  state-of-the-art lattice-regularized QCD calculations this work aims
at determining chiral pseudo-critical temperatures in QCD at zero and non-zero
chemical potentials \m{B,Q,S}, as well as for the situation analogous to the chemical
freeze-out stage of relativistic heavy-ion collision experiments. We will begin by
providing the necessary backgrounds in Sec.~\ref{sc:obs}, describe our methods in
Sec~\ref{sc:comp}, follow up with our results in Sec.~\ref{sc:results}, and end with
comparisons of our results with extant lattice QCD results and a short summary in
Sec.~\ref{sc:sum}.

\section{Observables and definitions}\label{sc:obs}

\subsection{Chiral observables}

To define the chiral order parameter we choose the combination
\begin{linenomath}
\begin{equation}\label{eq:pbp}
  \op = \frac{1}{f_K^4} \left[
  m_s \langle \qbq{u} + \qbq{d} \rangle
  - (m_u+m_d) \langle \qbq{s} \rangle \right] \,.
\end{equation}
\end{linenomath}
Here, \(\langle\qbq{q}\rangle=T(\partial{\ln Z}/\partial{m_f})/V\) denotes
chiral condensates of the up (\(u\)), down (\(d\)), and strange (\(s\)) quarks;
\(m_f\) denotes the masses of the quarks; \(Z\) is the partition function for
\(2+1\) flavor QCD, with \(m_u=m_d=m_s/27\), volume \(V\), temperature \(T\), and
\(\langle\cdot\rangle\) denotes  average over gauge configurations
corresponding to \(Z\). The susceptibility corresponding to the chiral order
parameter is defined as
\begin{linenomath}
  \begin{equation}  \label{eq:susg}
   \sust =   m_s\left( \frac{\partial}{\partial m_u} +
   \frac{\partial}{\partial m_d} \right) \op \,.
  \end{equation}
\end{linenomath}
\sust\ contains both quark-line connected, as well as quark-line disconnected pieces.
Since the singlet-axial \(U_A(1)\) symmetry of QCD is expected to remain broken at
all \(T\),  the quark-line connected piece is  expected to remain finite even for
\(m_u=m_d\to0\). Thus, we also separately consider the  quark-line disconnected
chiral susceptibility
\begin{linenomath}
  \begin{equation}\label{eq:sus}
   \sus = \frac{m_s^2}{f_K^4} \left[ \langle\left( \qbq{u} + \qbq{d} \right)^2\rangle
   - \left( \langle\qbq{u}\rangle + \langle\qbq{d}\rangle \right)^2 \right] \,.
  \end{equation}
\end{linenomath}
Note that, all chiral observables defined here are free of additive power
divergences and renormalization group invariant, ensuring existence of a continuum
limit up to small logarithmic corrections in \(m_f\). Additionally, all chiral
observables are defined to be dimensionless  in units of  the kaon decay constant
\(f_K=156.1/\sqrt{2}\)~MeV, the quantity used to determine lattice
spacing (\(a\))~\cite{Bazavov:2011nk}.

\subsection{Taylor expansions in chemical potentials}\label{sc:taylor}

The chemical potentials \m{u,d,s}\ of quarks in \(Z\) can be traded with the chemical
potentials \m{X}\ corresponding to any 3 other linearly independent conserved
charges, such as \m{B}, \m{S}, \m{Q} or \m{I}. Here, we choose to work with 2
independent sets \(\{B,Q,S\}\) and \(\{B,I,S\}\). \m{B,Q,S}\ are related to
\m{u,d,s}\ through \(\m{u}= \m{B}/3+2\m{Q}/3\), \(\m{d}=\m{B}/3-\m{Q}/3\), and
\(\m{s}=\m{B}/3-\m{Q}/3-\m{S}\). Similar relations for the \(\{B,I,S\}\) set are
\(\m{u}=\m{B}/3+\m{I}/2\), \(\m{d}=\m{B}/3-\m{I}/2\), and \(\m{s}=\m{B}/3-\m{S}\).

The \m{X} dependence of an observable, \emph{e.g.}, of \op, can be obtained by
following the well-established Taylor expansion method~\cite{Allton:2002zi,
Allton:2003vx, Gavai:2003mf} given by
\begin{linenomath}
  \begin{align}\label{eq:taylor}
    \begin{split}
      &\op(T,\m{X}) = \sum_{n=0}^{\infty} \frac{\opc{2n}(T)}{(2n)!}
      \left( \frac{\m{X}}{T} \right)^{2n} \,,
      \quad \mathrm{where} \quad \\
      &  \opc{2n}(T)  = \left.
      \frac{\partial^{2n} \op}{\partial \left( \m{X}/T \right)^{2n}}
      \right \vert_{\m{X}=0} \;.
    \end{split}
  \end{align}
\end{linenomath}
For simplicity, here we have assumed all  \(\m{Y\ne X}=0\). Due to
\(\mathcal{CP}\)-symmetry of \(Z\), Taylor expansions of  the chiral observables
contain only even powers of \m{X}. Similar expansions can be written for
\(\sus(T,\m{X})\), with Taylor coefficients \(\susc{2n}(T)\). For brevity,  we have
introduced the notations \(\opc{0}(T)=\op(T,0)\) and \(\susc{0}(T)=\sus(T,0)\). The
detailed expressions for \opc{2n}\ and \susc{2n}\ in terms of the \(u,d,s\) quark
propagators can be found in Refs.~\cite{Steinbrecher2018phd, Allton:2005gk}.

If \m{u,d,s}\ in \(Z\) are replaced by \m{B,Q,S}\ and, subsequently,
\(\m{Q}=\m{S}=0\) are imposed, then \m{B}\ will be given by the combination
\(\m{B}/3=\m{u}=\m{d}=\m{s}\). Exactly the same will happen for the \m{B,I,S}\ basis
if  \(\m{I}=\m{S}=0\) conditions are imposed. Similarly, for \(\m{B}=\m{Q}=0\) or
\(\m{B}=\m{I}=0\) both bases will lead to \(\m{S}=-\m{s}\), \(\m{u}=\m{d}=0\). Thus,
while computing the Taylor coefficients for \(B\) and \(S\) there is no need to
distinguish between \(\{B,Q,S\}\) and \(\{B,I,S\}\) bases. However, for
\(\m{B}=\m{S}=0\), in contrast to the \m{B,I,S}\ basis, Taylor coefficients with respect
to \m{Q}\ will receive additional contributions from the strange quark.

\subsection{Definitions of pseudo-critical temperatures}\label{sc:Tc-defs}

The nature of  the QCD chiral transition for \(m_u=m_d\to0\) and \(m_s>0\) remains an
open issue. Nevertheless, increasing numbers of sophisticated lattice QCD
calculations are now showing that, in this limit, the QCD chiral transition is most
likely a genuine second order phase transition that belongs to the 3D, \(O(4)\)
universality class~\cite{Ding:2018auz, Endrodi:2018xto, Burger:2011zc,
Cuteri:2018wci, Ejiri:2009ac}. On the other hand, for physical values of the quark masses
and vanishing chemical potentials, it is well established that chiral
symmetry restoration takes place via a smooth crossover~\cite{Bhattacharya:2014ara,
Bazavov:2011nk, Aoki:2006we}. The present work solely focuses on physical
\(m_{u,d,s}\). To ascribe precise meaning to chiral crossover temperatures we resort
to the well-defined notion of pseudo-critical temperatures \tc{X}.

In the vicinity of the second order chiral phase transition, behaviors of chiral
observables are governed by scaling properties of the 3D, \(O(4)\) universality
class~\cite{Ejiri:2009ac, Kaczmarek:2011zz}:
\begin{linenomath}
\begin{align}
\textstyle
\op(T,\m{B})\sim m^{1/\delta} f_G;\;\;
\sus(T,\m{B}), \sust(T) \sim m^{(1-\delta)/\delta} f_\chi
\end{align}
and
\begin{align}
\begin{split}
\partial_T \sust(T),\; \partial_T \susc{0}(T), \;&\susc{2}(T) \sim m^{(\beta-\beta\delta-1)/\beta\delta} f_\chi^\prime\,;\ \textstyle\\
\partial_T \opc{0}(T), \;&\opc{2}(T) \sim m^{(\beta-1)/\beta\delta}f_G^\prime\;. \textstyle
\end{split}
\end{align}
\end{linenomath}
Here, \opc{2n}\ and \susc{2n}\
are the coefficients of the Taylor series for \(\m{B}>0\) and \(\m{Q}=\m{S}=0\). The
two relevant scaling functions of the 3D, $O(4)$ universality class, \(f_G(z)\) and
\(f_\chi(z)\)~\cite{Engels:2011km,Engels:2014bra}, are functions of the so-called
scaling variable \(z=t/m^{1/\beta\delta}\), where \(m\sim m_{u,d}/m_s\),
\(t\sim(T-\ct)/\ct+K(\mu_B/T)^2\),  \ct\ is \(T_c(0)\) in the chiral limit
\(m\to0\), \(\beta\) and \(\delta\) are the critical exponents, and \(K\) is a
non-universal constant.

The chiral critical temperature \ct\ is defined as the temperature at which
\(\partial_T\op\) and \sust\ diverge in the limit \(V\to\infty\) and \(m\to0\).
For any \(m>0\), residing within the scaling regime, universality dictates that
\(\partial_T\op\) and \sust, scaled with appropriate (non-integer) powers of \(m\),
will have maxima located exactly at the maxima of the corresponding scaling functions
\(f_G^\prime(z)\)  and \(f_\chi(z)\). Thus, for \(m>0\) the locations of the maxima
of \(f_G^\prime(z)\) and \(f_\chi(z)\), denoted by \(z^G_p\) and  \(z^\chi_p\),
respectively, define  two pseudo-critical temperatures \(T_c^{G,\chi}(0)\). As
\(m\to0\), \(\partial_T\op\) and \sus\ diverge, and \(T_c^{G,\chi}(0)\) reduce to
\ct\ according to the scaling relation \(T_c^{G,\chi}(0) = \ct + A z^{G,\chi}_p
m^{1/\beta\delta}\), with a non-universal constant \(A\).

 Physical values of \(m_{u,d}\)  might not reside within the scaling regime of the
 second order chiral phase transition; consequently, chiral observables may also
 contain additional non-singular,  polynomial in \(m\), corrections. Thus, for
 physical values of \(m_{u,d}\)  we define \tco\ using the following criteria
 \begin{linenomath}
   \begin{align}\label{eq:tco-def}
     \begin{split}
       & \partial^2_T \opc{0}(T) = 0 \,,
       \quad
       \partial_T \opc{2}(T) = 0 \,, \\
       & \partial_T \sust(T) =0 \,,
       \quad
       \partial_T \susc{0}(T) = 0 \,,
       \quad
       \susc{2}(T) = 0 \,,
     \end{split}
   \end{align}
 \end{linenomath}
 where \opc{2n}\ and \susc{2n}\ are the coefficients of the Taylor series for
 \(\m{B}>0,~ \m{Q}=\m{S}=0\). Each of these 5 criteria may lead to 5 different values
 of  \tco, all of which will reduce to the unique \ct\ as \(m\to0\).  If the physical
 values of \(m_{u,d}\) happen to be in the scaling regime, then all these 5 criteria
 will lead to only two values of pseudo-critical temperatures \(T_c^{G,\chi}(0)\).
 The above definitions exhaust all second order fluctuations of the chiral order
 parameter through which locations of the maxima of \(f_G^\prime\) and \(f_\chi\)
 can be determined.

 Following the spirit of Taylor expansions, \(\m{X}\) dependence of pseudo-critical
 temperatures, up to \(\mathcal{O}(\m{X}^4)\), can be written as
 \begin{linenomath}
   \begin{equation}\label{eq:tc-exp}
     \!\!\tc{X} = \tco \left[ 1 -
     \kt{X} \!\left( \frac{\m{X}}{\tco} \right)^2 \!-
     \kf{X} \left( \frac{\m{X}}{\tco} \right)^4
     \right] \! .
   \end{equation}
 \end{linenomath}
As we will see later in Sec.~\ref{sc:T0}, for \(\m{X}=0\), the \tco\ defined through
all 5 criteria listed in Eq.~(\ref{eq:tco-def}) actually lead to the same result,
within our errors, in the continuum limit. Thus, for \(\m{X}>0\) it is sufficient to
define \tc{X} by the 2 criteria
\begin{linenomath}
  \begin{equation}\label{eq:tc-def}
    \left. \partial^2_T \op(T,\m{X}) \right\vert_{\m{X}} = 0  \,,
    \quad
    \left. \partial_T \sus(T,\m{X}) \right\vert_{\m{X}} = 0  \,.
  \end{equation}
\end{linenomath}
 Expressions for \kt{X} and \kf{X} can be obtained by:  (i) Expanding
 \(\op(T,\m{X})\), \(\sus(T,\m{X})\) in \m{X}\ using Eq.~(\ref{eq:taylor}); (ii)
 Taylor expanding \opc{2n}, \susc{2n}\ in powers of \((\tc{X}-\tco)\); (iii)
 Expanding \((\tc{X}-\tco)\) using Eq.~(\ref{eq:tc-exp}), keeping terms up to
 \(\mathcal{O}(\m{X}^4)\); (iv) Taking \(\partial_T\)  at fixed \m{X} of the
 fully-expanded expression up to \(\mathcal{O}(\m{X}^4)\), and imposing
 Eq.~(\ref{eq:tc-def}) order-by-order in \m{B}.  Since all quantities are assumed to
 be analytic in \m{X} around \(\m{X}=0\), all expansions in \m{X} and taking
 \(\partial_T\) can be carried out in any order, as long as all terms contributing up
 to \(\mathcal{O}(\m{X}^4)\)  are systematically included at each step. \emph{E.g.},
 for \sus\ we obtained~\cite{Steinbrecher2018phd}
 \begin{linenomath}
   \begin{align}\label{eq:kappa}
     \begin{split}
       \kt{X} &= \frac{1}{ 2 \D{2}\susc{0} }
       \left[ \Done\susc{2} - 2\susc{2} \right] \;, \\
       \kf{X} &= \frac{1}{ 24\D{2}\susc{0} } \left[
        - 72\kt{X}\susc{2} - 4\susc{4} + \Done\susc{4} \right. \\
        &+ \left.
       12\kt{X} \left(  4\Done\susc{2} - \D{2}\susc{2} + \kt{X}\D{3}\susc{0}
       \right) \right] \;, \\
     \end{split}
   \end{align}
 \end{linenomath}
where \susc{2n}\  are the expansion coefficients of  \sus\ with respect to \m{X}, and
the expressions are to be evaluated at \(T=\tco\). Similar expressions can be
obtained for \op~\cite{Steinbrecher2018phd}.

The expression for our \kt{B}\ corresponding to the order parameter is different from
that used in Ref.~\cite{Bonati:2018nut}, where \tc{B}\ was defined through
temperature derivatives at constant \(\m{B}/T\), rather than at constant \m{B}. We have
checked that the numerical results using both definitions are same within our errors.


\begin{figure*}[!t]
\centering
\vspace{-0.20em}
\includegraphics[width=0.325\textwidth]{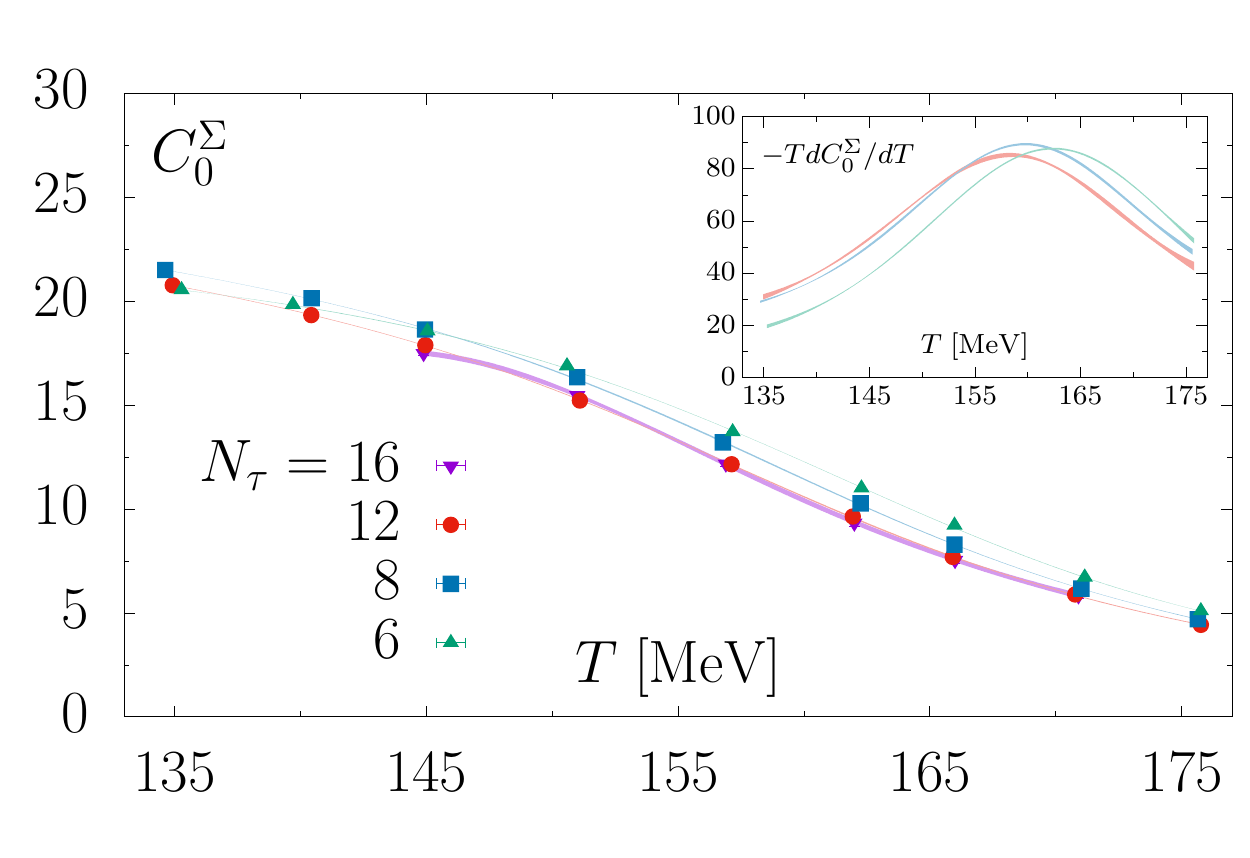}
\includegraphics[width=0.325\textwidth]{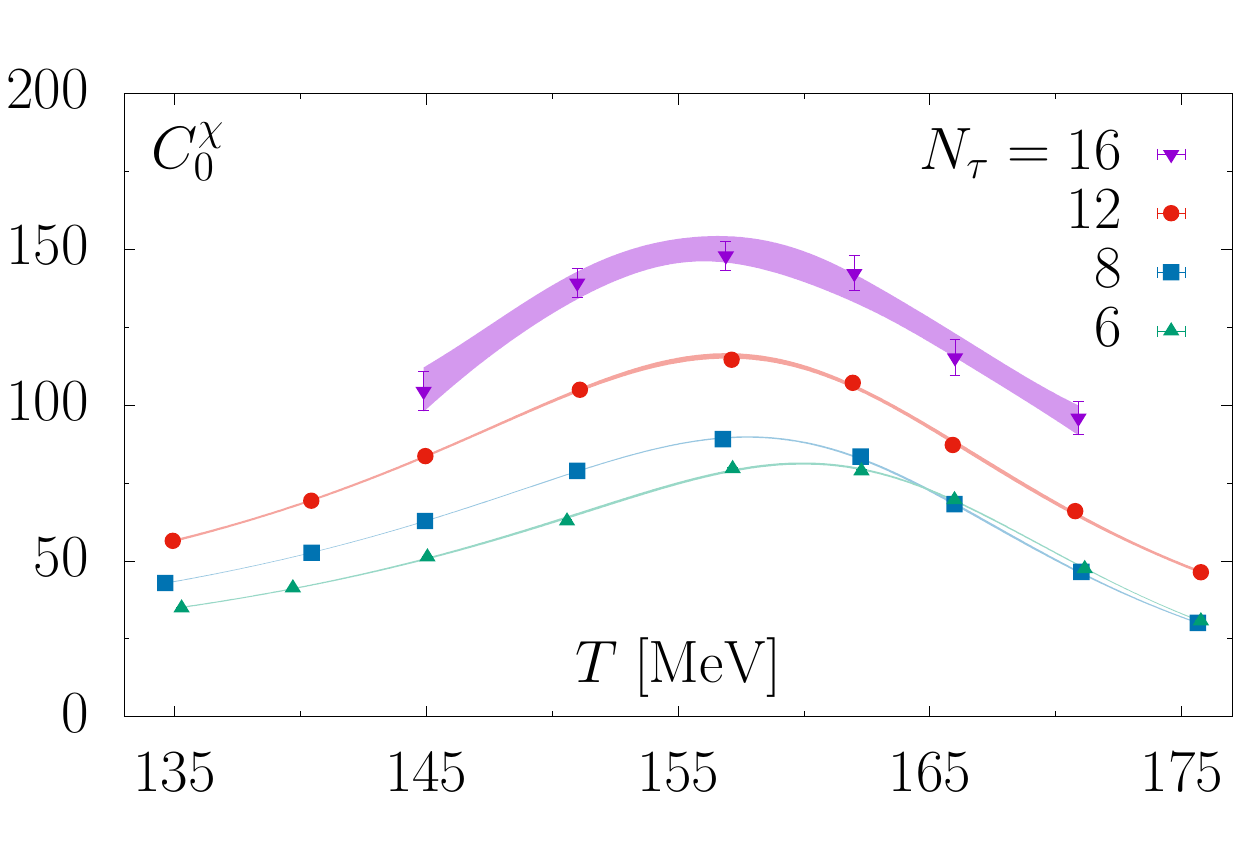}
\includegraphics[width=0.325\textwidth]{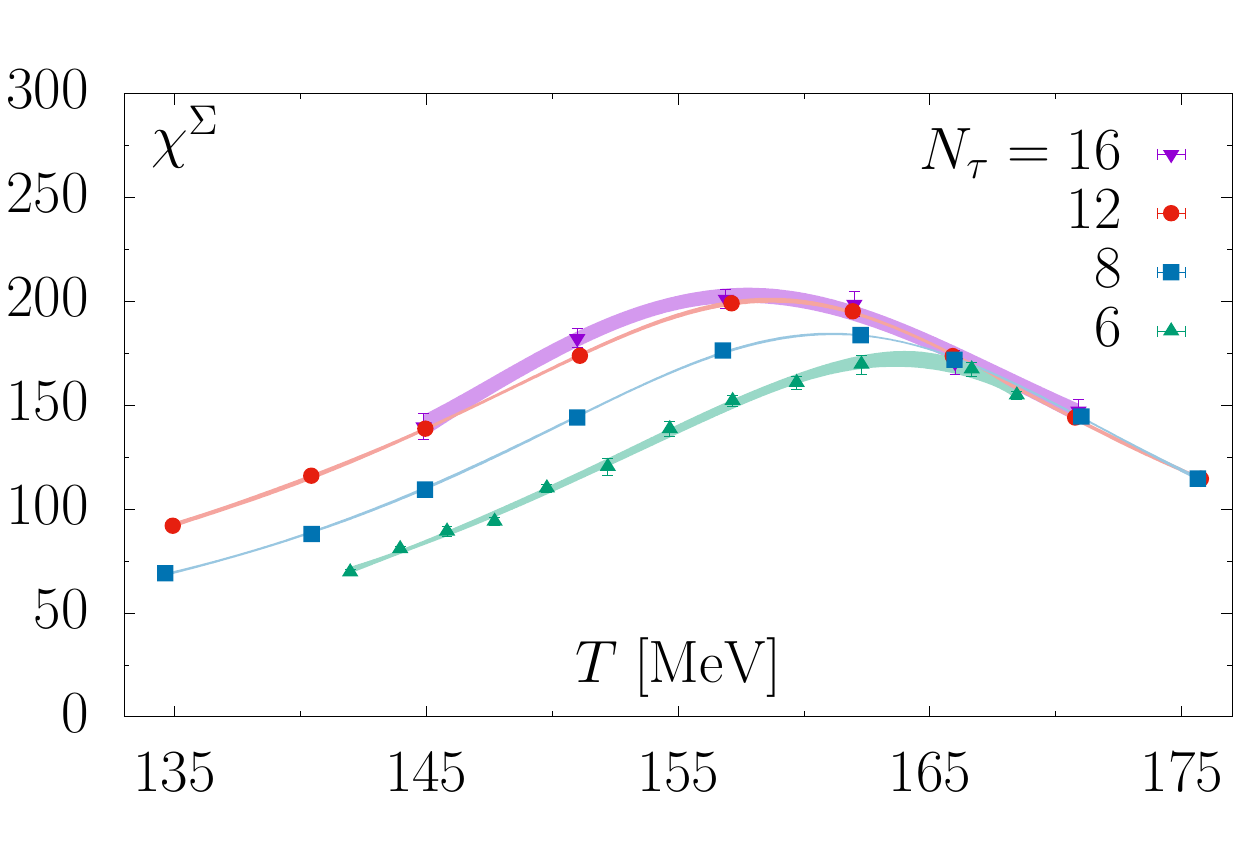}
\vspace{-1.25em}
\caption{Left: Chiral order parameter \(\opc{0}(T)=\op(T,\m{B,Q,S}=0)\). The inset
shows derivative of \opc{0}\ with respect to temperature \(T\). Middle: Disconnected
chiral susceptibility \(\susc{0}(T)\equiv\sus(T,\m{B,Q,S}=0)\). Right:
Susceptibility, \(\sust(T,\m{B,Q,S}=0)\), of the chiral order parameter.}
\vspace{-0.5em}
\label{fig:T0_1}
\end{figure*}

\section{Computational details}\label{sc:comp}

All computations presented in this study were carried out with the lattice actions
previously used by the HotQCD collaboration~\cite{Bazavov:2011nk, Bazavov:2012jq,
Bazavov:2017dus}, \emph{viz.}, the \(2+1\) flavor highly improved staggered
quarks (HISQ)~\cite{Follana:2006rc} and the tree-level improved Symanzik gauge
action.  The bare parameters of the lattice actions, \(m_u=m_d\), \(m_s\), and the
bare gauge coupling, are fixed by the line of constant physics determined by the
HotQCD collaboration~\cite{Bazavov:2011nk, Bazavov:2012jq, Bazavov:2017dus}. The
temperature is given by \(T=1/(a\nt)\), where \nt\ is the extent of the lattices
along the Euclidean temporal direction. The extents of the lattices along all 3
spatial directions were always chosen to be 4\nt, and the temporal extents were
varied from \nt=6, 8, 12, and 16, going towards progressively finer lattice spacing
at a fixed \(T\). Bare quark masses were chosen to reproduce, within a few percent,
the physical value of the kaon mass and a pseudo-Goldstone pion mass of 138~MeV in
the continuum limit at vanishing temperature and chemical potentials.

The fermionic operators needed to construct \opc{4}\ and \susc{2,4}\ were obtained
using the so-called linear-\m{}\ formalism~\cite{Gavai:2011uk, Gavai:2014lia,
Bazavov:2017dus}, but the traditional exponential-\m{}\
formalism~\cite{Hasenfratz:1983ba} was used for \opc{2}. On dimensional grounds,
within the linear-\m{} formalism no additive ultraviolet divergence (or constant) is
expected  in \susc{2n} for all \(n\),  and in \opc{2n}\ for
\(n>1\)~\cite{Steinbrecher2018phd}. To confirm these theoretical expectations we
computed \susc{2}\ by employing both linear- and exponential-\m{}\ formalism and
found identical results for both cases~\cite{Steinbrecher2018phd}. The
\(n^\mathrm{th}\) order Taylor coefficients of the chiral observables contain up to
\(n+1\) quark propagators, compared to \(n\) quark propagators for that in case of
the pressure (\(TV^{-1}\ln Z\)). Hence, the computational cost of \opc{2n}\ and
\susc{2n}\ increases accordingly.

All fermionic operators needed to construct the chiral observables and their Taylor
coefficients were measured on about 100K, 500K, 100K and 4K gauge field
configurations for \nt=6, 8, 12 and 16 lattices, respectively. In each case, the gauge
field configurations were separated by 10 rational hybrid Monte-Carlo trajectories of
unit length. The fermionic operators were calculated using the standard stochastic
estimator technique; more details about these computations can be found in
Ref.~\cite{Steinbrecher2018phd}.

As discussed in Sec.~\ref{sc:Tc-defs}, determinations of  \tco, \kt{X}\ and \kf{X}\
involve computing derivatives of the basic chiral observables and their Taylor
coefficients with respect to the temperature. To compute these derivatives, we
interpolated the basic observables in \(T\) between the computed data via the
following procedure. For each observable several \([m,n]\)  Pad\'e approximants were
used for \(N\) (\(>m+n\)) computed data, and \(N\) was varied by leaving out data
away from the crossover region. Statistical error of each Pad\'e approximant was
estimated using the bootstrap method; the bootstrap samples for each computed data
were drawn from a Gaussian distribution centered around the mean value of the data
and with a standard deviation equal to the \(1\sigma\) statistical error of that
data. The final \(T\)-interpolation for each observable was obtained by weighted
averaging over all the Pad\'e approximants where the weight for an approximant was
determined using the Akaike information criterion~\cite{Akaike:1974,
CAVANAUGH1997201}. This procedure gave reliable results for all the required
\(T\)-derivatives, especially for \(T\) in the vicinity of the
chiral-crossover~\cite{Steinbrecher2018phd}.

\begin{figure}[!t]
\centering
\vspace{-0.1em}
\includegraphics[width=0.465\textwidth]{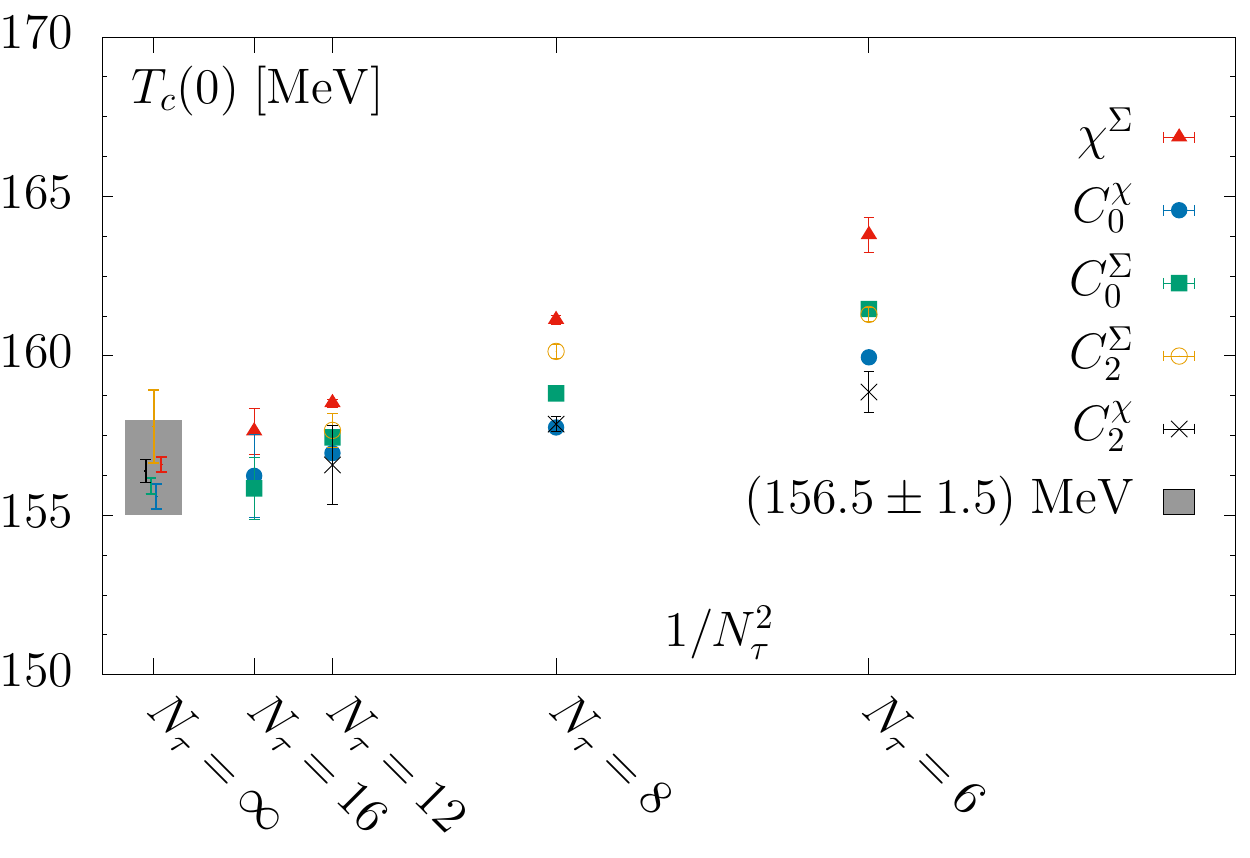}
\vspace{-0.9em}
\caption{Continuum extrapolations of pseudo-critical temperatures \(\tco\equiv
T_c(\m{B,Q,S}=0)\), defined using criteria listed in Eq.~(\ref{eq:tco-def}). The
solid gray band depicts the continuum-extrapolated result
\(\tco=(156.5\pm1.5)\)~MeV (see text for details).}
\vspace{-0.5em}
\label{fig:T0}
\end{figure}

We assumed that for all observables the leading discretization errors are of the type
\(a^2\propto1/\nt^2\). Extrapolations to the continuum limit \(a\to0\) were carried
out by fitting data at different \nt\ to a function linear in \(1/\nt^2\) and
extrapolating it to \(\nt\to\infty\) limit. The error on each continuum-extrapolated
result was obtained using the above described bootstrap method.  For all observables
we found that \(1/\nt^2\)-fits were satisfactory. To check the systematics of our
continuum extrapolations, we used fits including higher order
\(1/\nt^4\) corrections, as well as carried out the extrapolation procedure using an
alternative \(T\)-scale determined using the Sommer parameter \(r_1\); all results
were found to be consistent within our errors~\cite{Steinbrecher2018phd}.


\begin{figure*}[!t]
\centering
\includegraphics[width=0.325\textwidth]{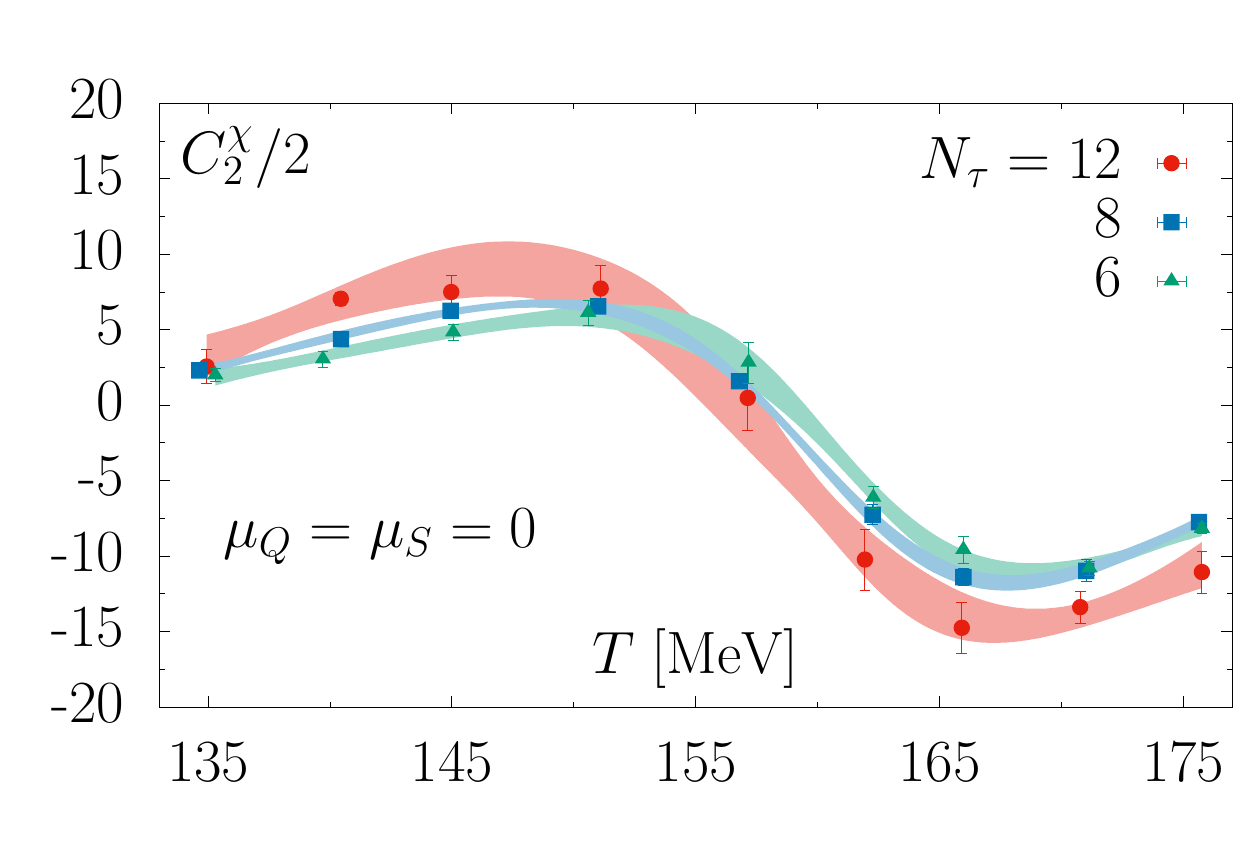}
\includegraphics[width=0.325\textwidth]{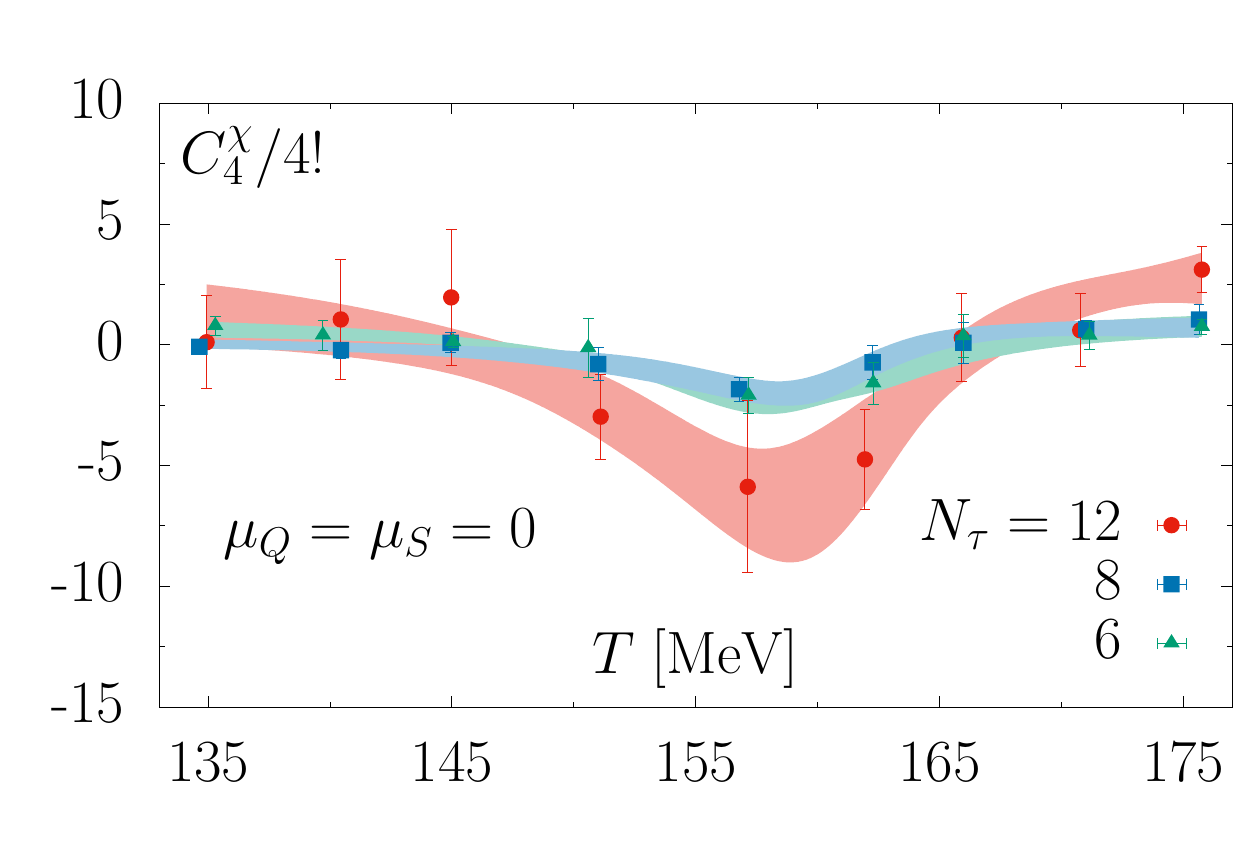}
\includegraphics[width=0.325\textwidth]{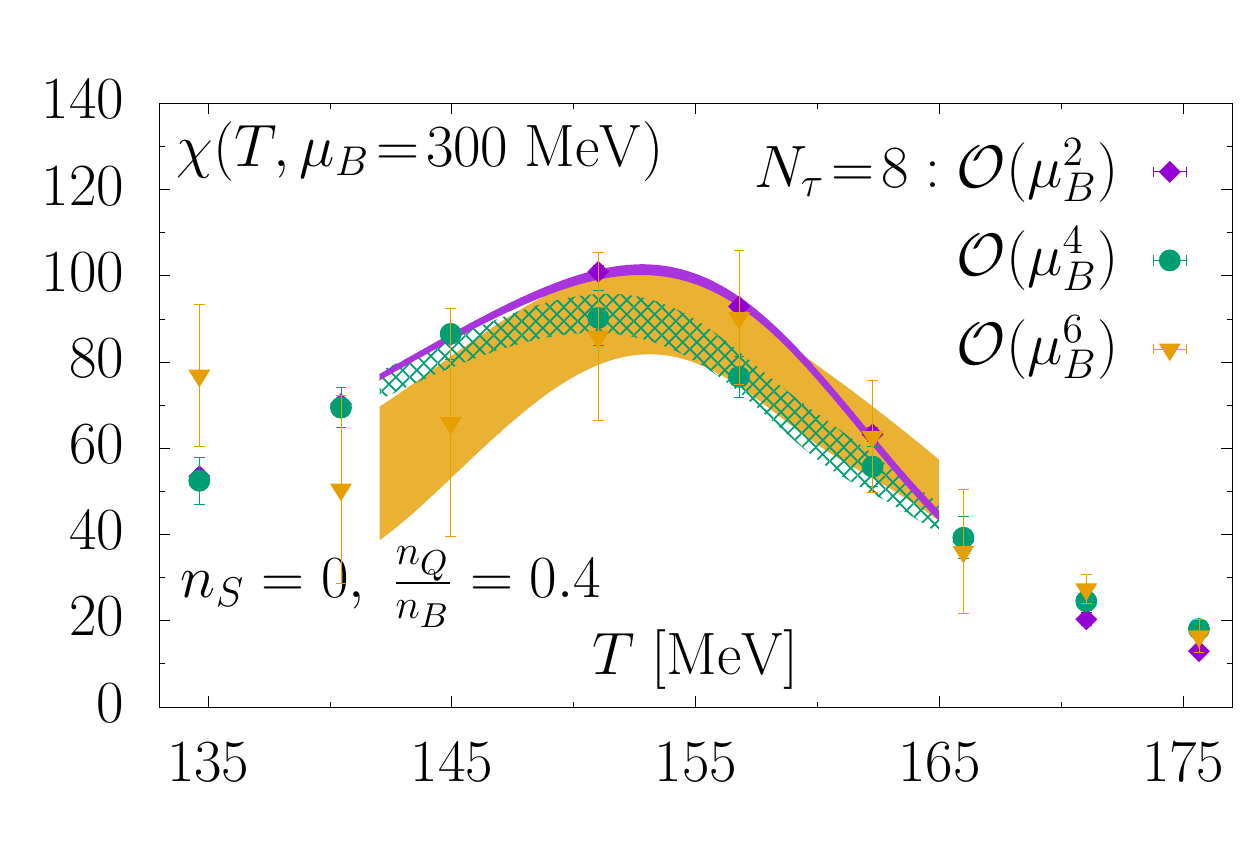}
\includegraphics[width=0.325\textwidth]{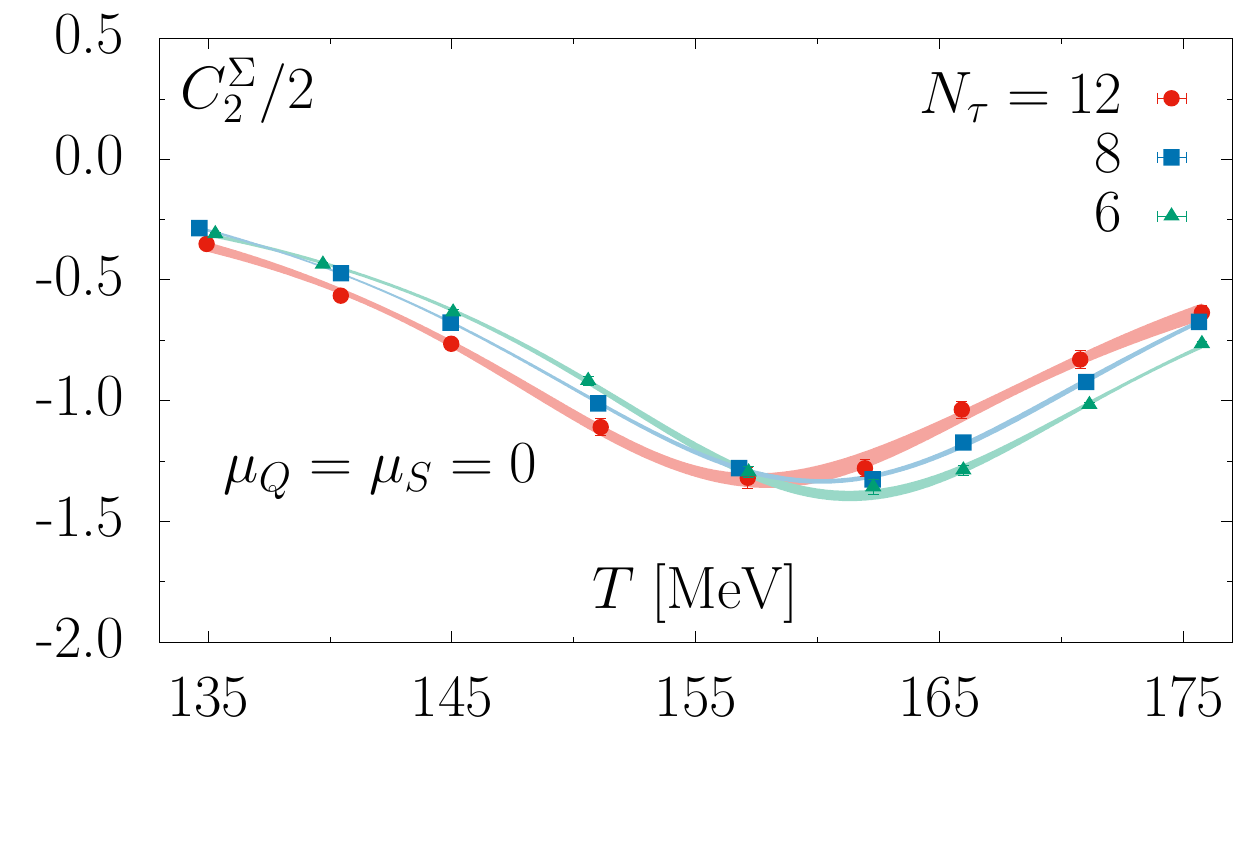}
\includegraphics[width=0.325\textwidth]{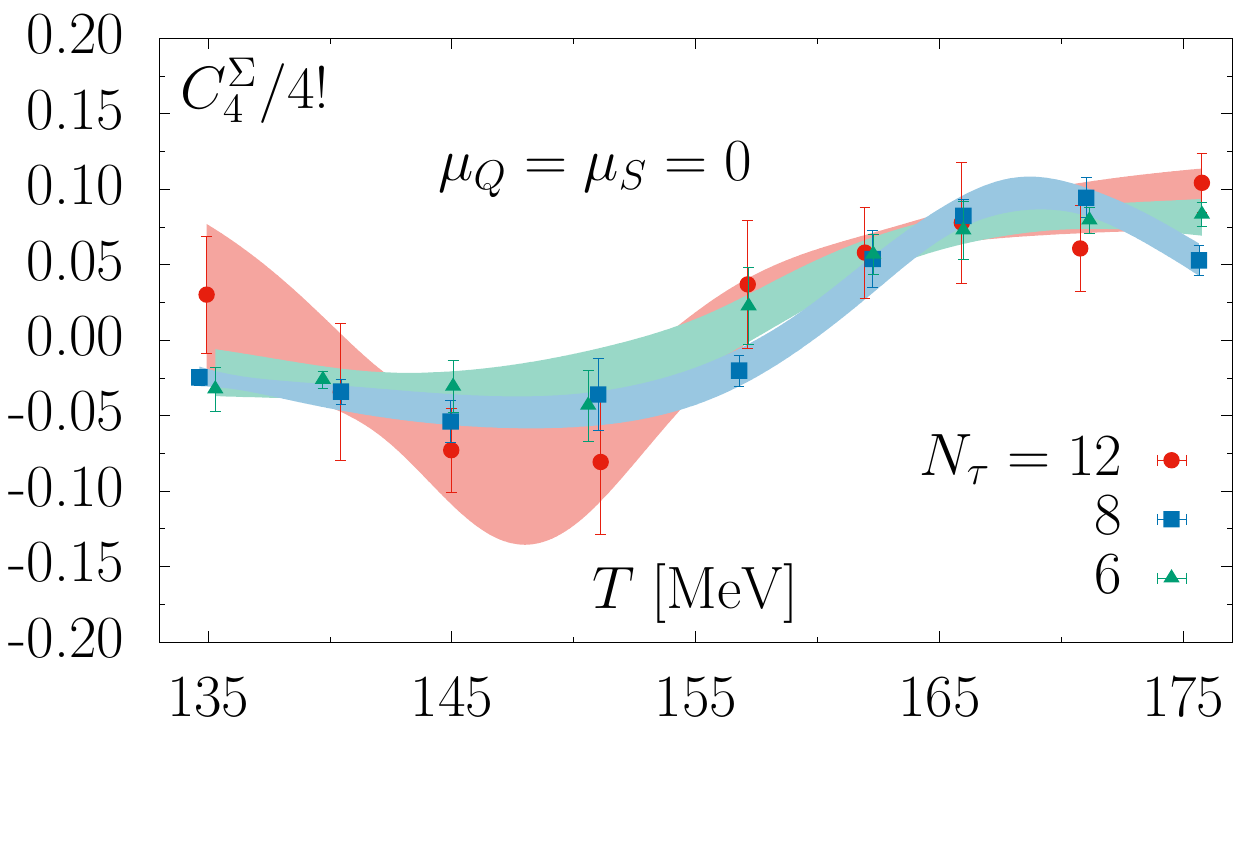}
\includegraphics[width=0.325\textwidth]{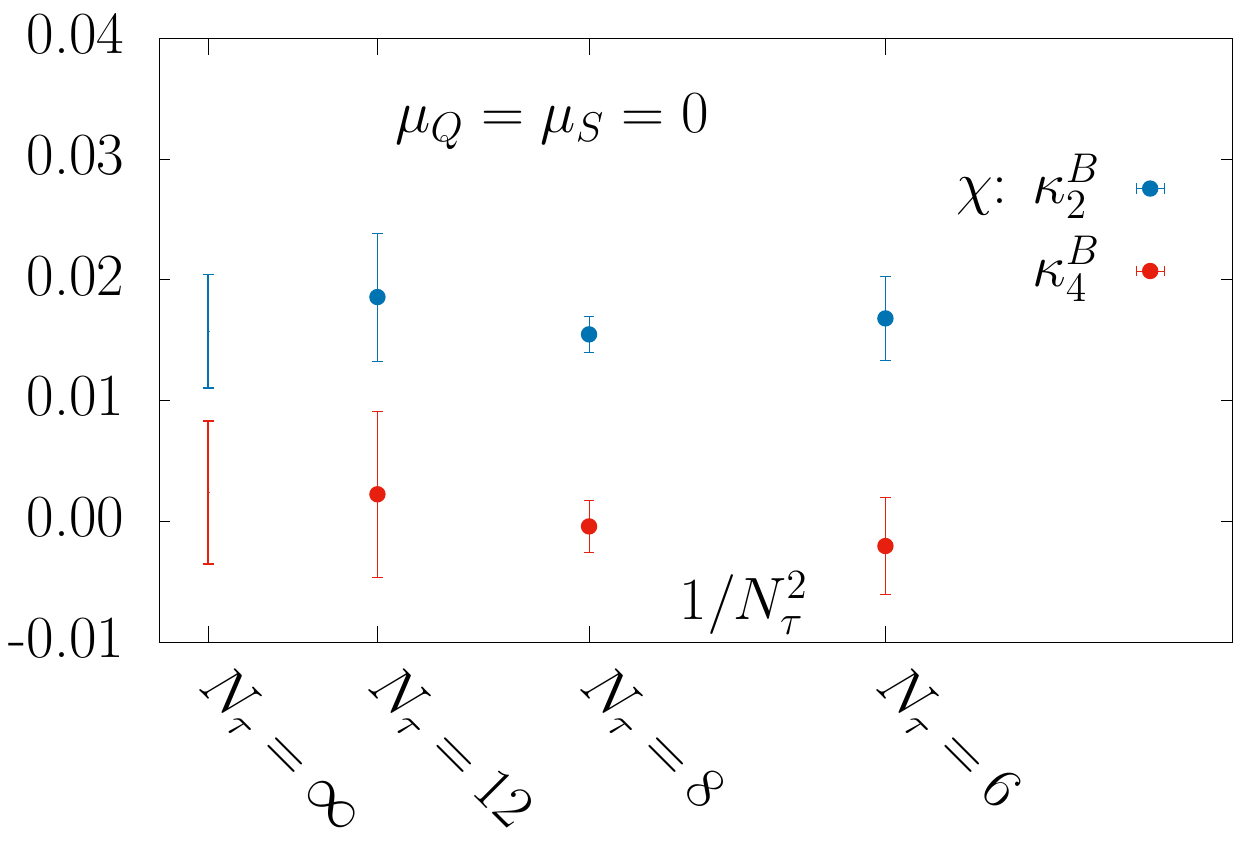}
\vspace{-0.5em}
\caption{Top-left: Second-order Taylor-coefficient \(\susc{2}(T)\), defined in
Eq.~(\ref{eq:taylor}), of the disconnected chiral susceptibility
\(\sus(T,\m{B},\m{Q}=\m{S}=0)\). Top-middle: Fourth-order Taylor-coefficient
\(\susc{4}(T)\) of \(\sus(T,\m{B},\m{Q}=\m{S}=0)\). Top-right: Order-by-order
corrections in \(\m{B}^{2n}\) to \(\sus(T,\m{B}=300~\mathrm{MeV},n_S=0,
n_Q=0.4n_B)\) for \nt=8\ lattices. Bottom-left: Second-order Taylor-coefficient
\(\opc{2}(T)\) of  the chiral order parameter \(\op(T,\m{B},\m{Q}=\m{S}=0)\).
Bottom-middle: Fourth-order Taylor-coefficient \(\opc{4}(T)\) of
\(\op(T,\m{B},\m{Q}=\m{S}=0)\).  Bottom-right: Second- (\kt{B}) and fourth-order
(\kf{B}) Taylor coefficients, defined in Eq.~(\ref{eq:tc-exp}), of the
pseudo-critical temperature \(T_c(\m{B},\m{Q}=\m{S}=0)\) obtained from
\(\sus(T,\m{B},\m{Q}=\m{S}=0)\).}

\label{fig:Tc}
\end{figure*}

\section{Results}\label{sc:results}

\subsection{Zero chemical potential: \tco}\label{sc:T0}

In Figs.~\ref{fig:T0_1} and \ref{fig:Tc}, we show all observables used for the
determination of pseudo-critical temperatures as defined in Eq.~(\ref{eq:tco-def})
for lattices with \nt=6, 8, 12, and 16. The results of the temperature
interpolations, obtained following the procedure described in Sec.~\ref{sc:comp}, are
shown by the corresponding solid bands. Using the interpolated results, and applying
the definitions in Eq.~(\ref{eq:tco-def}), we obtained 5 values of \tco\ for \nt=6,
8, and 12. These results are shown in Fig.~\ref{fig:T0}. Since we have not computed
\opc{2}\ and \susc{2}\ for \nt=16, we only show results for the other 3 definitions
of \tco. On coarser lattices, different definitions resulted in different values of
\tco. These differences progressively reduce with increasingly finer lattice spacing.
Results of \tco\ for each of the definitions were separately extrapolated to the
continuum (see Sec.~\ref{sc:comp} for details). The continuum-extrapolated results
for all 5 definitions of \tco\ were all consistent with each other within errors. We
took an unweighted average of all the 5 continuum results, and added the statistical
errors of each continuum-extrapolation in quadrature to quote our final result for
the chiral crossover temperature at zero chemical potentials
\(\tco=(156.5\pm1.5)\)~MeV. It is an interesting fact that continuum results for
different pseudo-critical temperatures coincide within a couple of MeV. However, if
the value of \(T_c^0\)~\cite{Ding:2018auz} is significantly different from \tco,
then, based on the scaling properties of \(T_c^{G,\chi}(0)\), it is natural to expect
more dispersion among the values of \tco. Coincidence of different pseudo-critical
temperatures for physical quark masses may accidentally arise due to the presence of
non-singular and/or sub-leading corrections to scaling. Further work will be needed
to clarify this issue.

\subsection{Non-zero chemical potentials: \kt{B,Q,S,I}\ and \kf{B,Q,S,I}}\label{sc:Tc}

Now, we present continuum-extrapolated results for the expansion coefficients \kt{X}\
and \kf{X}, defined by Eq.~(\ref{eq:tc-exp}), of \tc{X}\ for all conserved charges
\(X=B, S, Q, I\).   In all cases, extrapolations to the continuum were carried out
using results for \nt=6, 8, and 12. We discuss an example in detail, \emph{viz.},
\kt{B}\ and \kf{B} at \m{Q}=\m{S}=0. When \tc{B}\ is defined as the temperature where
\(\sus(T,\m{B})\) peaks at a given \m{B},  the corresponding \kt{B}\ and \kf{B}\ can
be obtained using Eq.~(\ref{eq:kappa}). The zeroth-, \(\susc{0}(T)\), second-,
\(\susc{2}(T)\), and the fourth-order, \(\susc{4}(T)\), expansion coefficients of
\(\sus(T,\m{B})\) in \(\m{B}/T\) (with \m{Q}=\m{S}=0) are shown in
Fig.~\ref{fig:T0_1} (middle), Fig.~\ref{fig:Tc} (top-left) and Fig.~\ref{fig:Tc}
(top-middle), respectively. The interpolations in \(T\) are shown by the
corresponding solid bands. Having determined \tco, \kt{B}\ and, subsequently, \kf{B}\
were obtained using the \(T\)-interpolations of \susc{0,2,4}. Similarly, \kt{B}\ and
\kf{B} were computed also from the inflection point of \(\op(T,\m{B})\) in \(T\), for
a given \m{B}, using the expansion coefficients \opc{0,2,4}, which are shown in
Fig.~\ref{fig:T0_1} (left), Fig.~\ref{fig:Tc} (bottom-left) and Fig.~\ref{fig:Tc}
(bottom-middle), respectively. Fig.~\ref{fig:Tc} (bottom-right) exemplifies the very
mild dependence of \kt{X}\ and \kf{X}\ on lattice spacing.

We also carried out similar computations to determine continuum-extrapolated \kt{X}\
and \kf{X}\ corresponding to  (i) \tc{S}\ at \m{B}=\m{Q}=0, (ii) \tc{Q}\ at
\m{B}=\m{S}=0, and (iii) \tc{I}\ at \m{B}=\m{S}=0; the values are listed in
Tab.~\ref{tb:kappa}. In all the cases, for both \kt{X}\ and \kf{X}, the results
obtained using two different definitions of \tc{X}, given in Eq.~(\ref{eq:tc-def}),
gave the same result within our errors. In each case, we took unweighted averages of
continuum-extrapolated results corresponding to both definitions for \tc{X}, and
added the respective statistical errors in quadrature to arrive at the final values
for \kt{X} and \kf{X}; these final results also are listed in the third row of
Tab.~\ref{tb:kappa}. In all cases, \kf{X} were found to be zero within errors, with
central values about an order of magnitude smaller than the corresponding \kt{X}.
Also, \kt{Q,I} were found to be about a factor 2 larger compared to \kt{B,S}.

\begin{table*}[!t]
  \centering
  \resizebox{\textwidth}{!}{
  \begin{tabular}{l  c c | c c | c c | c c | c c }
    & \kt{B} & \kf{B} & \kt{S} & \kf{S} & \kt{Q} & \kf{Q} & \kt{I} & \kf{I} &
    \kt{B,f} & \kf{B,f} \\ \hline
    \op  & 0.015(4) & -0.001(3) & 0.018(3) & 0.001(3) & 0.027(4) & 0.004(5) &
    0.023(3) & 0.004(4) & 0.012(2) & 0.000(2) \\
    \sus & 0.016(5) & 0.002(6) & 0.015(4) & 0.007(5) & 0.031(4) & 0.011(9) &
    0.028(3) & 0.006(6) & 0.012(3) & 0.000(4) \\ \hline
    Average & 0.016(6) & 0.001(7) & 0.017(5) & 0.004(6) & 0.029(6) & 0.008(1) &
    0.026(4) & 0.005(7)  & 0.012(4) & 0.000(4)
  \end{tabular}
  }

  \caption{Continuum-extrapolated values of second- (\kt{X}) and fourth-order
  (\kf{X}) Taylor coefficients, defined in Eq.~(\ref{eq:tc-exp}), of pseudo-critical
  temperature \tc{X=B,Q,S,I}\ obtained from the chiral order parameter \(\op(T,\m{X})\)
  and the disconnected chiral susceptibility  \(\sus(T,\m{X})\). Also listed are the
  continuum-extrapolated values of \kt{B,f}\ and \kf{B,f}\ for thermal conditions
  resembling the freeze-out stage of relativistic heavy-ion collisions, \emph{i.e.},
  \(\m{Q}(T,\m{B})\) and \(\m{S}(T,\m{B})\) fixed by strangeness-neutrality and
  isospin-imbalance of the colliding heavy-ions. The last row is obtained from
  unweighted average of the first two rows.}

\label{tb:kappa}
\end{table*}

\subsection{Heavy-ion collisions: \(\kappa_{2,4}^{B,f}\) for \(n_S=0\), \(n_Q=0.4n_B\)}\label{sc:HI}

In this case, \emph{i.e.}, for the thermal condition resembling the chemical
freeze-out stage of heavy-ion collision experiments,  we introduce the notations
\(\kappa_n^{B,f}\) as the Taylor coefficients of the corresponding pseudo-critical
temperature \(T_c^f(\m{B})\).

The formalism for Taylor expanding an observable in \(\m{B}/T\), with the
constraints \(n_S=0\) and \(n_Q=0.4n_B\), was introduced in
Ref.~\cite{Bazavov:2012vg} and has been applied to various
cases~\cite{Bazavov:2015zja, Bazavov:2017dus, Bazavov:2017tot}. With these
constraints, \m{S}\ and \m{Q}\ are no longer arbitrary, but become functions of  \(T\)
and \(\m{B}\). Following Ref.~\cite{Bazavov:2012vg},
\(\m{S}(T,\m{B})/T=s_1(T)\m{B}/T+s_3(\m{B}/T)^3\) and
\(\m{Q}(T,\m{B})/T=q_1(T)\m{B}/T+q_3(\m{B}/T)^3\) were Taylor-expanded in
\(\m{B}/T\). Expanding \(n_{B,Q,S}\) in powers of  \(\m{B}^i\m{Q}^j\m{S}^k\)
(\(i+j+k\le3\)), substituting expansions for \(\m{Q,S}(T,\m{B})\) in expansions of
\(n_{B,Q,S}\), and imposing the constraints \(n_S=0\) and \(n_Q=0.4n_B\)
order-by-order in \m{B}, expressions for \(s_{1,3}(T)\) and \(q_{1,3}(T)\) were
obtained in terms of the Taylor coefficients of the pressure. Explicit expressions
for \(s_{1,3}(T)\) and \(q_{1,3}(T)\)  can be found in Ref.~\cite{Bazavov:2017dus}.
By Taylor expanding \(\op(T,\m{B},\m{Q},\m{S})\) (\(\sus(T,\m{B},\m{Q},\m{S})\))  in powers
of \(\m{B}^i\m{Q}^j\m{S}^k\) (\(i+j+k\le4\)) and by using the expansions for
\(\m{Q,S}(T,\m{B})\), we obtained the expansions for \(\op(T,\m{B})\)
(\(\sus(T,\m{B})\)) up to \(\mathcal{O}(\m{B}^4)\).  As before, by invoking
Eq.~(\ref{eq:tc-def}), expressions were obtained for \(\kappa_{2,4}^{B,f}\).

Continuum-extrapolated results for \kt{B,f}\ and \kf{B,f}\ are given in
Tab.~\ref{tb:kappa}. \kt{B,f}\ came out to be same as \kt{B}\ and \kt{S}\ within
errors, and \kf{B,f}\ was found to be consistent with zero. On our \nt=8 lattices,
where we analyzed  half a million gauge configurations at all \(T\), we also computed
\(\m{B}^6\) corrections to the chiral observables. The order-by-order \m{B}\
corrections to \sus\ are shown in Fig.~\ref{fig:Tc} (top-right) at \m{B}=300 MeV and
for \(n_S=0\), \(n_Q=0.4n_B\). In the vicinity of \(T_c^{f}(\m{B})\),  difference
between \(\m{B}^4\) and  \(\m{B}^2\) corrections are clearly significant; but
\(\m{B}^6\) and \(\m{B}^4\) corrections are consistent within our errors. This shows
that up to \(\m{B}^4\) the expansion of \(T_c^{f}(\m{B})\) is controlled till
\(\m{B}\lesssim2\tco\). The phase boundary of QCD for \(n_S=0\), \(n_Q=0.4n_B\) is
shown in Fig.~\ref{fig:phase}; also shown are the chemical freeze-out points
extracted from heavy-ion collision experiments at various collision
energies~\cite{Andronic:2017pug, Adamczyk:2017iwn}, the line of constant energy
density
\(\epsilon(T,\m{B})=\epsilon(\tco,0)=0.42(6)~\mathrm{GeV/fm}^3\)~\cite{Bazavov:2017dus},
and the line of constant entropy density
\(s(T,\m{B})=s(\tco,0)=3.7(5)~\mathrm{fm}^{-3}\)~\cite{Bazavov:2017dus}.


\section{Discussions and summary}\label{sc:sum}

The value of  \tco\ reported in this work compares quite well with the previous
results from the HotQCD collaborations~\cite{Bazavov:2011nk, Bhattacharya:2014ara},
but the present result is about 6 times more accurate than the previous
continuum-extrapolated result~\cite{Bazavov:2011nk}.  Compared to that of
Ref.~\cite{Bazavov:2011nk}, use of 100-500 times more gauge configurations for \(\nt=
6, 8, 12\) in the present study resulted in the 6 times more accurate determination
of the continuum-extrapolated \tco. Our present value of \tco\  also is compatible
with the chiral pseudo-critical temperatures reported by other
groups~\cite{Borsanyi:2010bp, Bonati:2015bha}. It is pertinent to note that all our
calculations were carried out within a finite-size box of about 5 fm\(^3\) in the
vicinity of \tco; finite-size corrections might increase the value of \tco\ by an
amount commensurate to our present error on that quantity~\cite{HotQCD:chiralTc}.
\kt{B}\ determined in the present work is about a factor 2 larger than that reported
previously in Ref.~\cite{Endrodi:2011gv}.
\begin{figure}[!h]
\centering
\vspace{-2.5ex}
\includegraphics[width=0.465\textwidth]{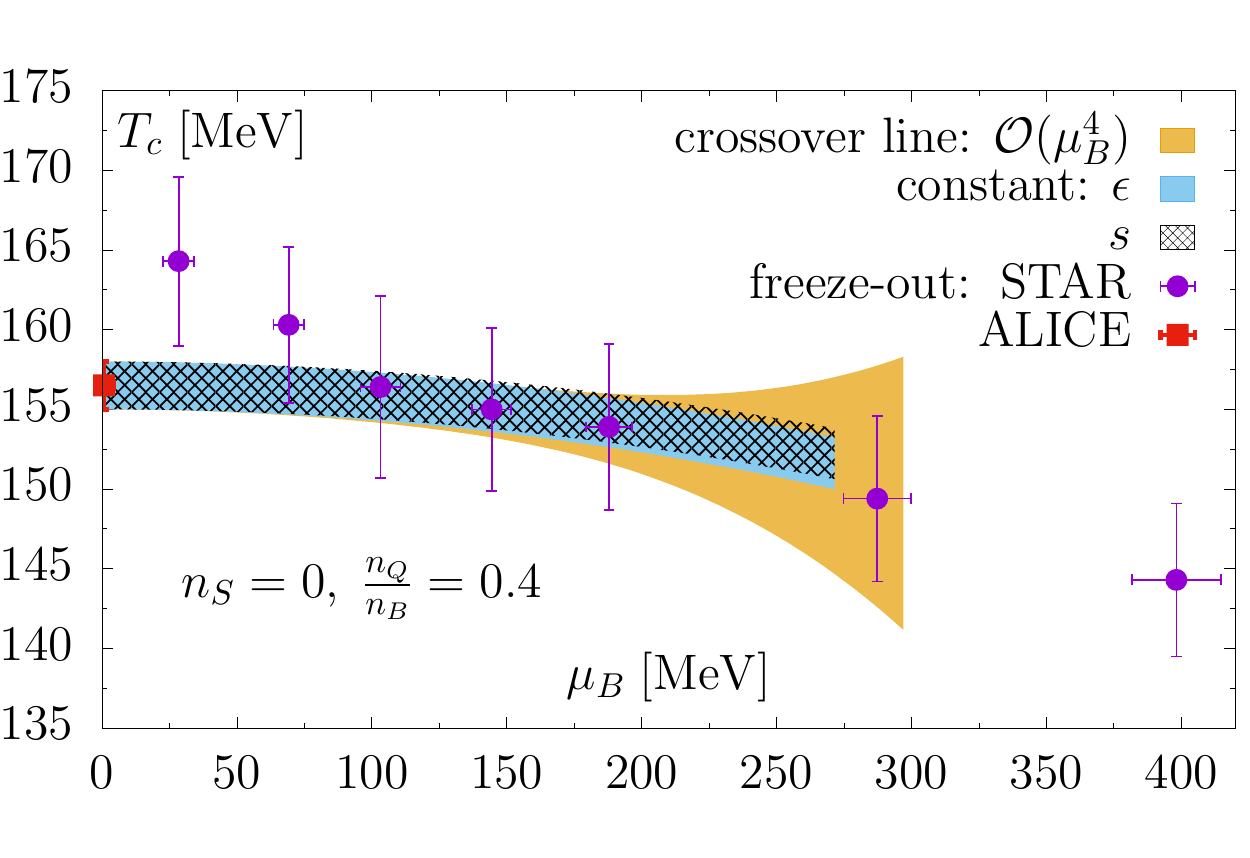}
\vspace{-2.5ex}

\caption{The phase boundary of \(2+1\) flavor QCD, with the constraints  \(n_S=0\)
and \(n_Q=0.4n_B\), is compared with the line of constant energy density
\(\epsilon=0.42(6)~\mathrm{GeV/fm}^3\) and the line of constant entropy density
\(s=3.7(5)~\mathrm{fm}^{-3}\)~\cite{Bazavov:2017dus} in the \(T\)-\(\m{B}\) plane.
Also, shown are the chemical freeze-out parameters extracted from grand canonical
ensemble based fits to hadron yields within 0-10\% centrality class for the
ALICE~\cite{Andronic:2017pug} experiment and 0-5\% centrality class for the
STAR~\cite{Adamczyk:2017iwn} experiment.}

\label{fig:phase}
\end{figure}
 Our present value of \kt{B}\ also is about
a factor 2 larger than the \kt{B}\ estimated using the curvature of the chiral
critical temperature along the light quark chemical potential
directions~\cite{Kaczmarek:2011zz}, but is consistent, within errors, with the same
reported in Ref.~\cite{Hegde:2015tbn}. In contrast to Ref.~\cite{Kaczmarek:2011zz},
Ref.~\cite{Hegde:2015tbn} used the much improved HISQ discretization. This clearly
suggests that the discrepancy between the present result and that estimated from
Ref.~\cite{Kaczmarek:2011zz} arises mostly due to the use of improved HISQ
discretization in the present study. On the other hand, \kt{B}\ reported in this work
is, within errors, compatible with those obtained in more recent works of
Refs.~\cite{Cea:2014xva, Bonati:2014rfa, Bonati:2015bha, Cea:2015cya}, obtained from
analytic continuations from purely imaginary \m{B}. It is also similar with that
obtained in Ref.~\cite{Bonati:2018nut} from Taylor expansion of chiral order
parameter for \(\m{B}>0\), \m{Q}=0 and \m{S}=\m{B}/3, in contrast to our choice of
\(\m{B}>0\) and \m{Q}=\m{S}=0. Our value of \kt{B,f}\ is quite similar to  that
reported in Ref.~\cite{Bellwied:2015rza}, determined from analytic continuations from
purely imaginary \m{}. Moreover, the phase boundary in the $T$-$\m{I}$ plane that can
be obtained using our \(\kappa_{2,4}^I\) is quite similar to that determined in
Ref.~\cite{Brandt:2017oyy} from lattice QCD computations performed directly at
\(\m{I}>0\), \m{B}=\m{S}=0.

In summary, using state-of-the-art lattice QCD computations we have determined
pseudo-critical temperatures,
\(\tc{X}=\tco[1-\kt{X}(\m{X}/\tco)^2-\kf{X}(\m{X}/\tco)^4]\), of QCD chiral crossover
for 6 different scenarios: (i) \tco\ for \(\m{B}=\m{Q}=\m{S}=0\); (ii)
\(\kappa_{2,4}^B\) for \(\m{B}>0\), \m{Q}=\m{S}=0; (iii) \(\kappa_{2,4}^S\) for
\(\m{S}>0\), \m{B}=\m{Q}=0; (iv) \(\kappa_{2,4}^Q\) for \(\m{Q}>0\), \m{B}=\m{S}=0;
(v) \(\kappa_{2,4}^I\) for \(\m{I}>0\), \m{B}=\m{S}=0; (vi) \(\kappa_{2,4}^{B,f}\)
for thermal conditions resembling that at the chemical freeze-out of relativistic
heavy-ion collision experiments, \emph{viz,} for \(\m{B}>0\), \(n_S=0\),
\(n_Q=0.4n_B\).  We have found
\begin{linenomath}
  \begin{equation}
    \tco = ( 156.5 \pm 1.5 )\ \mathrm{MeV} \,,
  \end{equation}
\end{linenomath}
and the values of \(\kappa_{2,4}^X\) are listed in Tab.~\ref{tb:kappa}. The QCD phase
boundary relevant for relativistic heavy-ion collision experiments have been
summarized in Fig.~\ref{fig:phase}. For \(\m{B}\lesssim300\)~MeV, the chemical
freeze-out takes place close to the QCD chiral crossover, which, in turn, seems to
happen along lines of constant energy density of \(0.42(6)~\mathrm{GeV/fm}^3\) and a
constant entropy density of \(3.7(5)~\mathrm{fm}^{-3}\).
At vanishing baryon chemical potential \(\m{B}\), the ALICE result~\cite{Andronic:2017pug} for the
chemical freeze-out temperature is in agreement with \tco. For \(\m{B}\lesssim300\)~MeV, all STAR
results~\cite{Adamczyk:2017iwn}, except the highest collision-energy, agree with \(\tc{B}\)
within their 1-sigma errors. The STAR result for the chemical freeze-out temperature at the
highest collision-energy agrees with \(\tc{B}\) within 1.5-sigma error. Thus, there
is no discrepancy between \(\tc{B}\) and chemical freeze-out temperatures extracted using
statistical model based fits to the experimentally measured hadron yields. However, it may pose a
challenge to the statistical hadronization based chemical freeze-out scenario if future improved
experiments determine freeze-out temperatures with statistical significance above \(\tc{B}\).

\vspace{-0.4em}
\section*{Acknowledgments}
\vspace{-0.2em}

This material is based upon work supported by the U.S. Department of Energy, Office of Science, Office of Nuclear Physics: (i) Through the Contract No. DE-SC0012704; (ii) Within the framework of the Beam Energy Scan Theory (BEST) Topical Collaboration; (iii) Through the Scientific Discovery through Advance Computing (ScIDAC) award Computing the Properties of Matter with Leadership Computing
Resources.

This research also was funded by--- (i) The Deutsche Forschungsgemeinschaft (DFG, German Research Foundation) - Project number 315477589-TRR 211; (ii) The grant 05P18PBCA1 of the German Bundesministerium f\"ur Bildung und Forschung; (iii) The National Natural Science Foundation of China under grant numbers 11775096 and 11535012 (HTD); (iv) The Early Career Research Award of the Science and Engineering Research Board of the Government of India (PH); (v)  Ramanujan  Fellowship  of  the  Department  of  Science  and Technology,  Government  of  India (SS).

This research used awards of computer time provided by the INCITE and ALCC programs at:  (i) Oak Ridge Leadership Computing Facility, a DOE Office of Science User Facility operated under Contract No. DE-AC05-00OR22725; (ii) National Energy Research Scientific Computing Center, a U.S. Department of Energy Office of Science User Facility operated under Contract No. DE-AC02-05CH11231; (iii) Argonne Leadership Computing Facility, a U.S. Department of Energy Office of Science User Facility operated under Contract No. DE-AC02-06CH11357.

This research also used computing resources made available through:  (i) The USQCD resources at BNL, FNAL and JLAB; (ii) The  PRACE grants at CSCS, Switzerland, and CINECA, Italy; (iii) The Gauss Center at NIC-J\"ulich, Germany; (iv) Nuclear Science Computing Center at Central China Normal University.

\bibliographystyle{elsarticle-num}
\bibliography{ref}

\begin{thebibliography}{10}
\expandafter\ifx\csname url\endcsname\relax
  \def\url#1{\texttt{#1}}\fi
\expandafter\ifx\csname urlprefix\endcsname\relax\def\urlprefix{URL }\fi
\expandafter\ifx\csname href\endcsname\relax
  \def\href#1#2{#2} \def\path#1{#1}\fi

\bibitem{Ding:2015ona}
H.-T. Ding, F.~Karsch, S.~Mukherjee, {Thermodynamics of strong-interaction
  matter from Lattice QCD}, Int. J. Mod. Phys. E24~(10) (2015) 1530007.
\newblock \href {http://arxiv.org/abs/1504.05274} {\path{arXiv:1504.05274}},
  \href {https://doi.org/10.1142/S0218301315300076}
  {\path{doi:10.1142/S0218301315300076}}.

\bibitem{Wygas:2018otj}
M.~M. Wygas, I.~M. Oldengott, {D. B\"odeker}, D.~J. Schwarz, {Cosmic QCD Epoch
  at Nonvanishing Lepton Asymmetry}, Phys. Rev. Lett. 121~(20) (2018) 201302.
\newblock \href {http://arxiv.org/abs/1807.10815} {\path{arXiv:1807.10815}},
  \href {https://doi.org/10.1103/PhysRevLett.121.201302}
  {\path{doi:10.1103/PhysRevLett.121.201302}}.

\bibitem{Fukushima:2010bq}
K.~Fukushima, T.~Hatsuda, {The phase diagram of dense QCD}, Rept. Prog. Phys.
  74 (2011) 014001.
\newblock \href {http://arxiv.org/abs/1005.4814} {\path{arXiv:1005.4814}},
  \href {https://doi.org/10.1088/0034-4885/74/1/014001}
  {\path{doi:10.1088/0034-4885/74/1/014001}}.

\bibitem{Busza:2018rrf}
W.~Busza, K.~Rajagopal, W.~van~der Schee, {Heavy Ion Collisions: The Big
  Picture, and the Big Questions}, Ann. Rev. Nucl. Part. Sci. 68 (2018)
  339--376.
\newblock \href {http://arxiv.org/abs/1802.04801} {\path{arXiv:1802.04801}},
  \href {https://doi.org/10.1146/annurev-nucl-101917-020852}
  {\path{doi:10.1146/annurev-nucl-101917-020852}}.

\bibitem{Andronic:2017pug}
A.~Andronic, P.~Braun-Munzinger, K.~Redlich, J.~Stachel, {Decoding the phase
  structure of QCD via particle production at high energy}, Nature 561~(7723)
  (2018) 321--330.
\newblock \href {http://arxiv.org/abs/1710.09425} {\path{arXiv:1710.09425}},
  \href {https://doi.org/10.1038/s41586-018-0491-6}
  {\path{doi:10.1038/s41586-018-0491-6}}.

\bibitem{Bazavov:2011nk}
A.~Bazavov, et~al., {The chiral and deconfinement aspects of the QCD
  transition}, Phys. Rev. D85 (2012) 054503.
\newblock \href {http://arxiv.org/abs/1111.1710} {\path{arXiv:1111.1710}},
  \href {https://doi.org/10.1103/PhysRevD.85.054503}
  {\path{doi:10.1103/PhysRevD.85.054503}}.

\bibitem{Allton:2002zi}
C.~R. Allton, S.~Ejiri, S.~J. Hands, O.~Kaczmarek, F.~Karsch, E.~Laermann,
  C.~Schmidt, L.~Scorzato, {The QCD thermal phase transition in the presence of
  a small chemical potential}, Phys. Rev. D66 (2002) 074507.
\newblock \href {http://arxiv.org/abs/hep-lat/0204010}
  {\path{arXiv:hep-lat/0204010}}, \href
  {https://doi.org/10.1103/PhysRevD.66.074507}
  {\path{doi:10.1103/PhysRevD.66.074507}}.

\bibitem{Allton:2003vx}
C.~R. Allton, S.~Ejiri, S.~J. Hands, O.~Kaczmarek, F.~Karsch, E.~Laermann,
  C.~Schmidt, {The Equation of state for two flavor QCD at nonzero chemical
  potential}, Phys. Rev. D68 (2003) 014507.
\newblock \href {http://arxiv.org/abs/hep-lat/0305007}
  {\path{arXiv:hep-lat/0305007}}, \href
  {https://doi.org/10.1103/PhysRevD.68.014507}
  {\path{doi:10.1103/PhysRevD.68.014507}}.

\bibitem{Gavai:2003mf}
R.~V. Gavai, S.~Gupta, {Pressure and nonlinear susceptibilities in QCD at
  finite chemical potentials}, Phys. Rev. D68 (2003) 034506.
\newblock \href {http://arxiv.org/abs/hep-lat/0303013}
  {\path{arXiv:hep-lat/0303013}}, \href
  {https://doi.org/10.1103/PhysRevD.68.034506}
  {\path{doi:10.1103/PhysRevD.68.034506}}.

\bibitem{Steinbrecher2018phd}
P.~Steinbrecher, \href{https://pub.uni-bielefeld.de/record/2919977}{{The QCD
  crossover up to \(\mathcal{O}(\mu_B^6)\) from Lattice QCD}}, Ph.D. thesis,
  Universit\"at Bielefeld (2018).
\newline\urlprefix\url{https://pub.uni-bielefeld.de/record/2919977}

\bibitem{Allton:2005gk}
C.~R. Allton, M.~Doring, S.~Ejiri, S.~J. Hands, O.~Kaczmarek, F.~Karsch,
  E.~Laermann, K.~Redlich, {Thermodynamics of two flavor QCD to sixth order in
  quark chemical potential}, Phys. Rev. D71 (2005) 054508.
\newblock \href {http://arxiv.org/abs/hep-lat/0501030}
  {\path{arXiv:hep-lat/0501030}}, \href
  {https://doi.org/10.1103/PhysRevD.71.054508}
  {\path{doi:10.1103/PhysRevD.71.054508}}.

\bibitem{Ding:2018auz}
  H.-T.~Ding, P.~Hegde, F.~Karsch, A.~Lahiri, S.-T.~Li, S.~Mukherjee and P.~Petreczky,
  {Chiral phase transition of (2+1)-flavor QCD,}
  Nucl.\ Phys.\ A {\bf 982} (2019) 211,
  doi:10.1016/j.nuclphysa.2018.10.032,
\newblock \href {http://arxiv.org/abs/1807.05727} {\path{arXiv:1807.05727}},
\newblock \href {http://arxiv.org/abs/1905.11610} {\path{arXiv:1905.11610}}.


\bibitem{Endrodi:2018xto}
G.~Endrodi, L.~Gonglach, {Chiral transition via the Banks-Casher relation}, in:
  {36th International Symposium on Lattice Field Theory (Lattice 2018) East
  Lansing, MI, United States, July 22-28, 2018}, 2018.
\newblock \href {http://arxiv.org/abs/1810.09173} {\path{arXiv:1810.09173}}.

\bibitem{Burger:2011zc}
F.~Burger, E.-M. Ilgenfritz, M.~Kirchner, M.~P. Lombardo, {M.
  M\"uller-Preussker}, O.~Philipsen, C.~Urbach, L.~Zeidlewicz, {Thermal QCD
  transition with two flavors of twisted mass fermions}, Phys. Rev. D87~(7)
  (2013) 074508.
\newblock \href {http://arxiv.org/abs/1102.4530} {\path{arXiv:1102.4530}},
  \href {https://doi.org/10.1103/PhysRevD.87.074508}
  {\path{doi:10.1103/PhysRevD.87.074508}}.

\bibitem{Cuteri:2018wci}
F.~Cuteri, O.~Philipsen, A.~Sciarra, {Progress on the nature of the QCD thermal
  transition as a function of quark flavors and masses}, 2018.
\newblock \href {http://arxiv.org/abs/1811.03840} {\path{arXiv:1811.03840}}.

\bibitem{Ejiri:2009ac}
S.~Ejiri, F.~Karsch, E.~Laermann, C.~Miao, S.~Mukherjee, P.~Petreczky,
  C.~Schmidt, W.~Soeldner, W.~Unger, {On the magnetic equation of state in
  (2+1)-flavor QCD}, Phys. Rev. D80 (2009) 094505.
\newblock \href {http://arxiv.org/abs/0909.5122} {\path{arXiv:0909.5122}},
  \href {https://doi.org/10.1103/PhysRevD.80.094505}
  {\path{doi:10.1103/PhysRevD.80.094505}}.

\bibitem{Bhattacharya:2014ara}
T.~Bhattacharya, et~al., {QCD Phase Transition with Chiral Quarks and Physical
  Quark Masses}, Phys. Rev. Lett. 113~(8) (2014) 082001.
\newblock \href {http://arxiv.org/abs/1402.5175} {\path{arXiv:1402.5175}},
  \href {https://doi.org/10.1103/PhysRevLett.113.082001}
  {\path{doi:10.1103/PhysRevLett.113.082001}}.

\bibitem{Aoki:2006we}
Y.~Aoki, G.~Endrodi, Z.~Fodor, S.~D. Katz, K.~K. Szabo, {The Order of the
  quantum chromodynamics transition predicted by the standard model of particle
  physics}, Nature 443 (2006) 675.
\newblock \href {http://arxiv.org/abs/hep-lat/0611014}
  {\path{arXiv:hep-lat/0611014}}, \href {https://doi.org/10.1038/nature05120}
  {\path{doi:10.1038/nature05120}}.

\bibitem{Kaczmarek:2011zz}
O.~Kaczmarek, F.~Karsch, E.~Laermann, C.~Miao, S.~Mukherjee, P.~Petreczky,
  C.~Schmidt, W.~Soeldner, W.~Unger, {Phase boundary for the chiral transition
  in (2+1) -flavor QCD at small values of the chemical potential}, Phys. Rev.
  D83 (2011) 014504.
\newblock \href {http://arxiv.org/abs/1011.3130} {\path{arXiv:1011.3130}},
  \href {https://doi.org/10.1103/PhysRevD.83.014504}
  {\path{doi:10.1103/PhysRevD.83.014504}}.

\bibitem{Engels:2011km}
J.~Engels, F.~Karsch, {The scaling functions of the free energy density and its
  derivatives for the 3d O(4) model}, Phys. Rev. D85 (2012) 094506.
\newblock \href {http://arxiv.org/abs/1105.0584} {\path{arXiv:1105.0584}},
  \href {https://doi.org/10.1103/PhysRevD.85.094506}
  {\path{doi:10.1103/PhysRevD.85.094506}}.

\bibitem{Engels:2014bra}
J.~Engels, F.~Karsch, {Finite size dependence of scaling functions of the
  three-dimensional O(4) model in an external field}, Phys. Rev. D90~(1) (2014)
  014501.
\newblock \href {http://arxiv.org/abs/1402.5302} {\path{arXiv:1402.5302}},
  \href {https://doi.org/10.1103/PhysRevD.90.014501}
  {\path{doi:10.1103/PhysRevD.90.014501}}.

\bibitem{Bonati:2018nut}
C.~Bonati, M.~D'Elia, F.~Negro, F.~Sanfilippo, K.~Zambello, {Curvature of the
  pseudocritical line in QCD: Taylor expansion matches analytic continuation},
  Phys. Rev. D98~(5) (2018) 054510.
\newblock \href {http://arxiv.org/abs/1805.02960} {\path{arXiv:1805.02960}},
  \href {https://doi.org/10.1103/PhysRevD.98.054510}
  {\path{doi:10.1103/PhysRevD.98.054510}}.

\bibitem{Bazavov:2012jq}
A.~Bazavov, et~al., {Fluctuations and Correlations of net baryon number,
  electric charge, and strangeness: A comparison of lattice QCD results with
  the hadron resonance gas model}, Phys. Rev. D86 (2012) 034509.
\newblock \href {http://arxiv.org/abs/1203.0784} {\path{arXiv:1203.0784}},
  \href {https://doi.org/10.1103/PhysRevD.86.034509}
  {\path{doi:10.1103/PhysRevD.86.034509}}.

\bibitem{Bazavov:2017dus}
A.~Bazavov, et~al., {The QCD Equation of State to $\mathcal{O}(\mu_B^6)$ from
  Lattice QCD}, Phys. Rev. D95~(5) (2017) 054504.
\newblock \href {http://arxiv.org/abs/1701.04325} {\path{arXiv:1701.04325}},
  \href {https://doi.org/10.1103/PhysRevD.95.054504}
  {\path{doi:10.1103/PhysRevD.95.054504}}.

\bibitem{Follana:2006rc}
E.~Follana, Q.~Mason, C.~Davies, K.~Hornbostel, G.~P. Lepage, J.~Shigemitsu,
  H.~Trottier, K.~Wong, {Highly improved staggered quarks on the lattice, with
  applications to charm physics}, Phys. Rev. D75 (2007) 054502.
\newblock \href {http://arxiv.org/abs/hep-lat/0610092}
  {\path{arXiv:hep-lat/0610092}}, \href
  {https://doi.org/10.1103/PhysRevD.75.054502}
  {\path{doi:10.1103/PhysRevD.75.054502}}.

\bibitem{Gavai:2011uk}
R.~V. Gavai, S.~Sharma, {A faster method of computation of lattice quark number
  susceptibilities}, Phys. Rev. D85 (2012) 054508.
\newblock \href {http://arxiv.org/abs/1112.5428} {\path{arXiv:1112.5428}},
  \href {https://doi.org/10.1103/PhysRevD.85.054508}
  {\path{doi:10.1103/PhysRevD.85.054508}}.

\bibitem{Gavai:2014lia}
R.~V. Gavai, S.~Sharma, {Divergences in the quark number susceptibility: The
  origin and a cure}, Phys. Lett. B749 (2015) 8--13.
\newblock \href {http://arxiv.org/abs/1406.0474} {\path{arXiv:1406.0474}},
  \href {https://doi.org/10.1016/j.physletb.2015.07.036}
  {\path{doi:10.1016/j.physletb.2015.07.036}}.

\bibitem{Hasenfratz:1983ba}
P.~Hasenfratz, F.~Karsch, {Chemical Potential on the Lattice}, Phys. Lett. 125B
  (1983) 308--310.
\newblock \href {https://doi.org/10.1016/0370-2693(83)91290-X}
  {\path{doi:10.1016/0370-2693(83)91290-X}}.

\bibitem{Akaike:1974}
H.~Akaike, A new look at the statistical model identification, IEEE
  Transactions on Automatic Control 19~(6) (1974) 716.
\newblock \href {https://doi.org/10.1109/TAC.1974.1100705}
  {\path{doi:10.1109/TAC.1974.1100705}}.

\bibitem{CAVANAUGH1997201}
J.~E. Cavanaugh, {Unifying the derivations for the Akaike and corrected Akaike
  information criteria}, Statistics \& Probability Letters 33~(2) (1997) 201.
\newblock \href {https://doi.org/10.1016/S0167-7152(96)00128-9}
  {\path{doi:10.1016/S0167-7152(96)00128-9}}.

\bibitem{Bazavov:2012vg}
A.~Bazavov, et~al., {Freeze-out Conditions in Heavy Ion Collisions from QCD
  Thermodynamics}, Phys. Rev. Lett. 109 (2012) 192302.
\newblock \href {http://arxiv.org/abs/1208.1220} {\path{arXiv:1208.1220}},
  \href {https://doi.org/10.1103/PhysRevLett.109.192302}
  {\path{doi:10.1103/PhysRevLett.109.192302}}.

\bibitem{Bazavov:2015zja}
A.~Bazavov, et~al., {Curvature of the freeze-out line in heavy ion collisions},
  Phys. Rev. D93~(1) (2016) 014512.
\newblock \href {http://arxiv.org/abs/1509.05786} {\path{arXiv:1509.05786}},
  \href {https://doi.org/10.1103/PhysRevD.93.014512}
  {\path{doi:10.1103/PhysRevD.93.014512}}.

\bibitem{Bazavov:2017tot}
A.~Bazavov, et~al., {Skewness and kurtosis of net baryon-number distributions
  at small values of the baryon chemical potential}, Phys. Rev. D96~(7) (2017)
  074510.
\newblock \href {http://arxiv.org/abs/1708.04897} {\path{arXiv:1708.04897}},
  \href {https://doi.org/10.1103/PhysRevD.96.074510}
  {\path{doi:10.1103/PhysRevD.96.074510}}.

\bibitem{Adamczyk:2017iwn}
L.~Adamczyk, et~al., {Bulk Properties of the Medium Produced in Relativistic
  Heavy-Ion Collisions from the Beam Energy Scan Program}, Phys. Rev. C96~(4)
  (2017) 044904.
\newblock \href {http://arxiv.org/abs/1701.07065} {\path{arXiv:1701.07065}},
  \href {https://doi.org/10.1103/PhysRevC.96.044904}
  {\path{doi:10.1103/PhysRevC.96.044904}}.

\bibitem{Borsanyi:2010bp}
S.~Borsanyi, Z.~Fodor, C.~Hoelbling, S.~D. Katz, S.~Krieg, C.~Ratti, K.~K.
  Szabo, {Is there still any $T_c$ mystery in lattice QCD? Results with
  physical masses in the continuum limit III}, JHEP 09 (2010) 073.
\newblock \href {http://arxiv.org/abs/1005.3508} {\path{arXiv:1005.3508}},
  \href {https://doi.org/10.1007/JHEP09(2010)073}
  {\path{doi:10.1007/JHEP09(2010)073}}.

\bibitem{Bonati:2015bha}
C.~Bonati, M.~D'Elia, M.~Mariti, M.~Mesiti, F.~Negro, F.~Sanfilippo, {Curvature
  of the chiral pseudocritical line in QCD: Continuum extrapolated results},
  Phys. Rev. D92~(5) (2015) 054503.
\newblock \href {http://arxiv.org/abs/1507.03571} {\path{arXiv:1507.03571}},
  \href {https://doi.org/10.1103/PhysRevD.92.054503}
  {\path{doi:10.1103/PhysRevD.92.054503}}.

\bibitem{HotQCD:chiralTc}
H.~T.~Ding {\it et al.},
  {The chiral phase transition temperature in (2+1)-flavor QCD,}
\newblock \href {http://arxiv.org/abs/1903.04801} {\path{arXiv:1903.04801}}.

\bibitem{Endrodi:2011gv}
G.~Endrodi, Z.~Fodor, S.~D. Katz, K.~K. Szabo, {The QCD phase diagram at
  nonzero quark density}, JHEP 04 (2011) 001.
\newblock \href {http://arxiv.org/abs/1102.1356} {\path{arXiv:1102.1356}},
  \href {https://doi.org/10.1007/JHEP04(2011)001}
  {\path{doi:10.1007/JHEP04(2011)001}}.

\bibitem{Hegde:2015tbn}
P.~Hegde, H.-T. Ding, {The curvature of the chiral phase transition line for
  small values of $\mu_B$}, PoS LATTICE2015 (2016) 141.
\newblock \href {http://arxiv.org/abs/1511.03378} {\path{arXiv:1511.03378}},
  \href {https://doi.org/10.22323/1.251.0141} {\path{doi:10.22323/1.251.0141}}.

\bibitem{Cea:2014xva}
P.~Cea, L.~Cosmai, A.~Papa, {Critical line of 2+1 flavor QCD}, Phys. Rev.
  D89~(7) (2014) 074512.
\newblock \href {http://arxiv.org/abs/1403.0821} {\path{arXiv:1403.0821}},
  \href {https://doi.org/10.1103/PhysRevD.89.074512}
  {\path{doi:10.1103/PhysRevD.89.074512}}.

\bibitem{Bonati:2014rfa}
C.~Bonati, M.~D'Elia, M.~Mariti, M.~Mesiti, F.~Negro, F.~Sanfilippo, {Curvature
  of the chiral pseudocritical line in QCD}, Phys. Rev. D90~(11) (2014) 114025.
\newblock \href {http://arxiv.org/abs/1410.5758} {\path{arXiv:1410.5758}},
  \href {https://doi.org/10.1103/PhysRevD.90.114025}
  {\path{doi:10.1103/PhysRevD.90.114025}}.

\bibitem{Cea:2015cya}
P.~Cea, L.~Cosmai, A.~Papa, {Critical line of 2+1 flavor QCD: Toward the
  continuum limit}, Phys. Rev. D93~(1) (2016) 014507.
\newblock \href {http://arxiv.org/abs/1508.07599} {\path{arXiv:1508.07599}},
  \href {https://doi.org/10.1103/PhysRevD.93.014507}
  {\path{doi:10.1103/PhysRevD.93.014507}}.

\bibitem{Bellwied:2015rza}
R.~Bellwied, S.~Borsanyi, Z.~Fodor, {G\"unther J.}, S.~D. Katz, C.~Ratti, K.~K.
  Szabo, {The QCD phase diagram from analytic continuation}, Phys. Lett. B751
  (2015) 559--564.
\newblock \href {http://arxiv.org/abs/1507.07510} {\path{arXiv:1507.07510}},
  \href {https://doi.org/10.1016/j.physletb.2015.11.011}
  {\path{doi:10.1016/j.physletb.2015.11.011}}.

\bibitem{Brandt:2017oyy}
B.~B. Brandt, G.~Endrodi, S.~Schmalzbauer, {QCD phase diagram for nonzero
  isospin-asymmetry}, Phys. Rev. D97~(5) (2018) 054514.
\newblock \href {http://arxiv.org/abs/1712.08190} {\path{arXiv:1712.08190}},
  \href {https://doi.org/10.1103/PhysRevD.97.054514}
  {\path{doi:10.1103/PhysRevD.97.054514}}.

\end{thebibliography}
\end{document}